\documentclass[journal,letterpaper]{IEEEtran_nssmic}
\usepackage{graphicx}  
\usepackage{amsmath}   
\usepackage{array}

\begin{document}

\def\d{\partial}
\def\um{\,\mu{\rm  m}}
\def\mm{\,   {\rm mm}}
\def\cm{\,   {\rm cm}}
\def \m{\,   {\rm  m}}
\def\ps{\,   {\rm ps}}
\def\ns{\,   {\rm ns}}
\def\us{\,\mu{\rm  s}}
\def\ms{\,   {\rm ms}}
\def\nA{\,   {\rm nA}}
\def\uA{\,\mu{\rm  A}}
\def\mA{\,   {\rm mA}}
\def\A {\,   {\rm  A}}
\def\mV{\,   {\rm mV}}
\def\V {\,   {\rm  V}}
\def\fF{\,   {\rm fF}}
\def\pF{\,   {\rm pF}}
\def\GeV{\, {\rm GeV}}
\def\MHz{\, {\rm MHz}}
\def\uW{\,\mu{\rm  W}}
\def\e {\,  {\rm e^-}}

\renewcommand{\labelenumi}{\arabic{enumi}}
\renewcommand{\labelitemi}{-}

\title{Trends in Pixel Detectors: Tracking and Imaging}
\author{Norbert~Wermes
\thanks{N. Wermes is with Bonn University, Physikalisches Institut, Nussallee 12, D-53115 Bonn, Germany,
email: wermes@physik.uni-bonn.de}%
\thanks{Work supported by the German Ministerium f{\"u}r Bildung,
              Wissenschaft, Forschung und Technologie (BMBF) under contract
              no.~$05 HA8PD1$, by the
              Ministerium f{\"u}r Wissenschaft und Forschung des Landes
              Nordrhein--Westfalen under contract no.~IV$\,A5-106\,011\,98$, and
              by the Deutsche Forschungsgemeinschaft DFG}
}
\specialpapernotice{(Invited Paper)}
\maketitle

\begin{abstract}
For large scale applications, hybrid pixel detectors, in which
sensor and read-out IC are separate entities, constitute the state
of the art in pixel detector technology to date. They have been
developed and start to be used as tracking detectors and also
imaging devices in radiography, autoradiography, protein
crystallography and in X-ray astronomy. A number of trends and
possibilities for future applications in these fields with
improved performance, less material, high read-out speed, large
radiation tolerance, and potential off-the-shelf availability have
appeared and are momentarily matured. Among them are monolithic or
semi-monolithic approaches which do not require complicated
hybridization but come as single sensor/IC entities. Most of these
are presently still in the development phase waiting to be used as
detectors in experiments. The present state in pixel detector
development including hybrid and (semi-)monolithic pixel
techniques and their suitability for particle detection and for
imaging, is reviewed.
\end{abstract}


%

%

\section{Introduction}
%
Albeit being similar at first sight, the requirements for charged
particle detection in high energy physics compared to imaging
applications with pixel detectors can be very different. In
particle physics experiments charged particles, usually triggered
by other subdetectors, have to be identified with high demands on
spatial resolution and timing. In imaging applications the image
is obtained by the un-triggered accumulation (integrating or
counting) of the quanta of the impinging radiation, often also
with high demands (e.g. $>$ 1 MHz per pixel in certain radiography
or CT applications). Si pixel detectors for high energy charge
particle detection can assume typical signal charges collected at
an electrode in the order of 5,000-10,000 electrons even taking
into account charge sharing between cells and detector
deterioration after irradiation to doses as high as $60$ Mrad. In
$^3$H-autoradiography, on the contrary, or in X-ray astronomy the
amount of charge to be collected with high efficiency can be much
below $1000$ e. The spatial resolution is governed by the
attainable pixel granularity from a few to about 10$\mu$m at best,
obtained with pixel dimensions in the order of $50 \mu$m to $100
\mu$m. The requirements from radiology are similar. Mammography
($\sim$ 80$\mu$m pixel dimensions) is most demanding with respect
to space resolution. Some applications in autoradiography,
however, require sub-$\mu$m resolutions, not attainable with
present day semiconductor detectors. For applications with lower
demands on the spatial resolution ($PSF \sim $5-10$\mu$m) but with
requirements on real time and time resolved data acquisition,
semiconductor pixel detectors are very attractive.

Very thin detector assemblies are mandatory for vertex detectors
at collider experiments, in particular for the planned linear
$e^+e^-$ collider. This imposes high demands on the detector
development. While silicon is almost a perfect material for
particle physics detectors, allowing the shaping of electric
fields by tailored impurity doping, the need of high photon
absorption efficiency in radiological applications requires the
study and use of semiconductor materials with high atomic charge,
such as GaAs or CdTe. For such materials the charge collection
properties are much less understood and mechanical issues in
particular those related to hybrid pixels are abundant, most
notably regarding the hybridization of detectors when they are not
available in wafer scale sizes. Finally, for applications in high
radiation environments, such as hadron colliders, radiation hard
sensors are developed using either Si with a dedicated design or
CVD-diamond.

\section{Hybrid Pixels: The State of the Art in Pixel Detector Technology}
\begin{figure}[h]
\begin{center}
\includegraphics[width=0.45\textwidth]{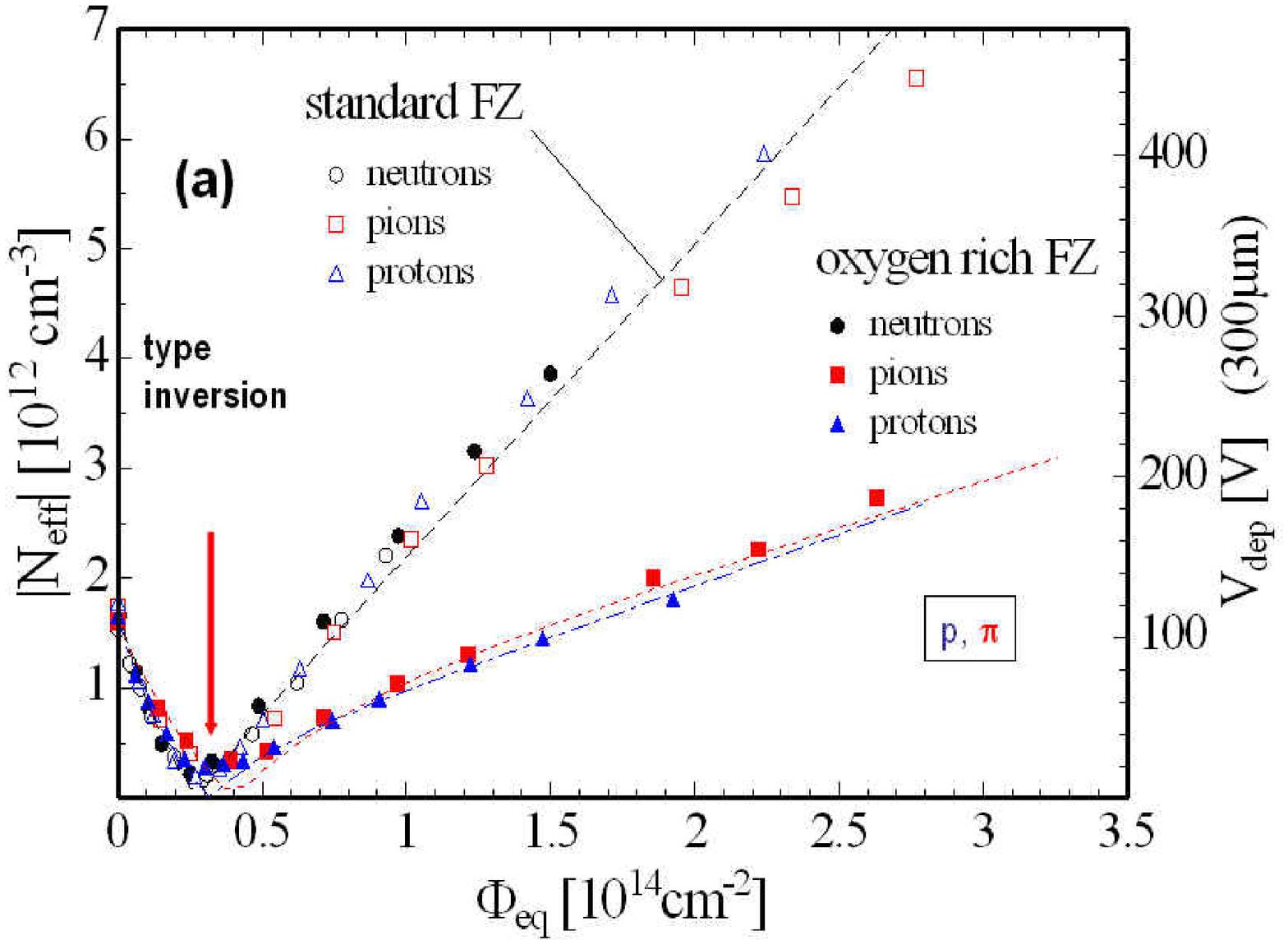}
\hfill \vspace{0.5cm} \break
\includegraphics[width=0.20\textwidth]{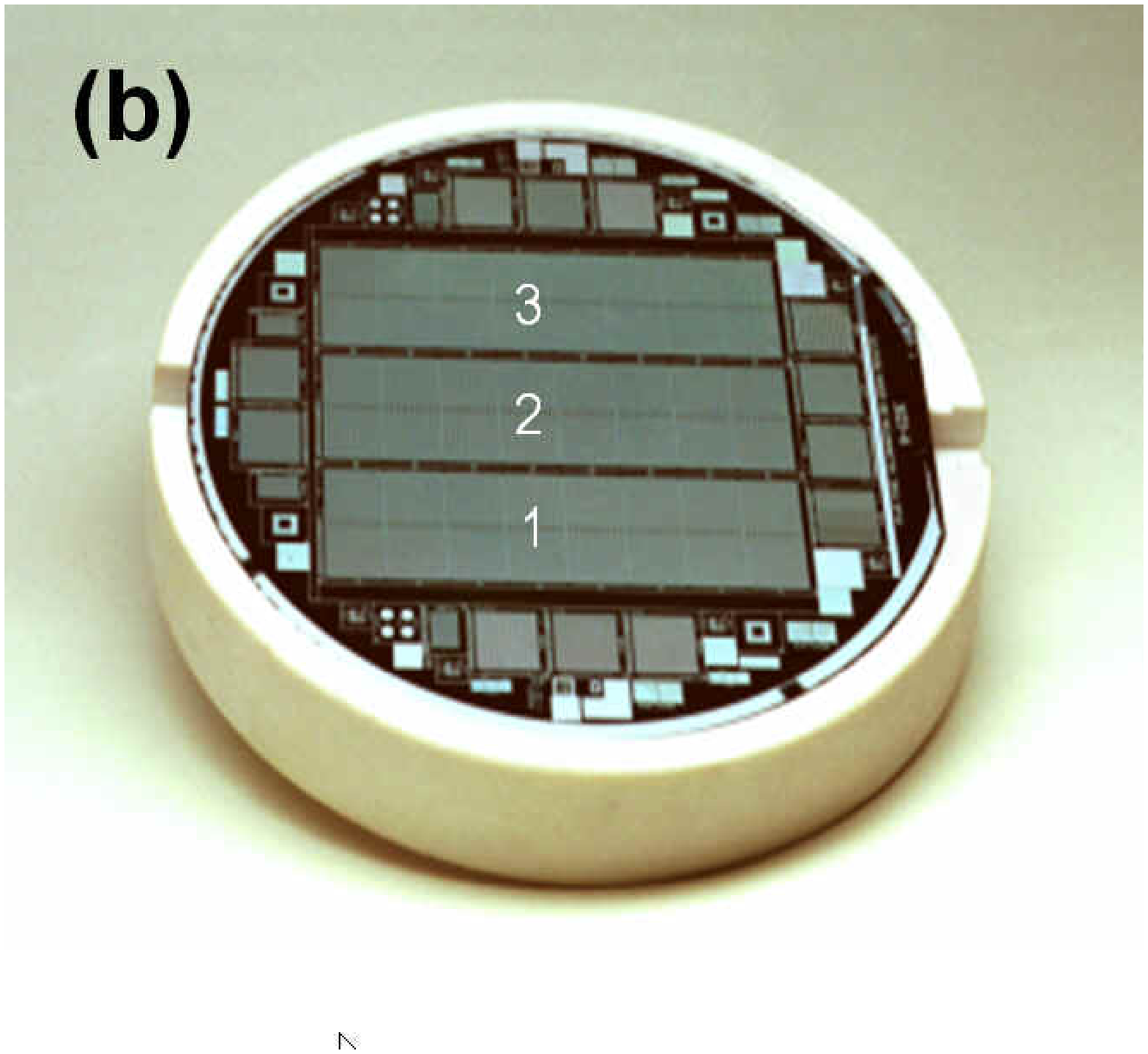}
\includegraphics[width=0.28\textwidth]{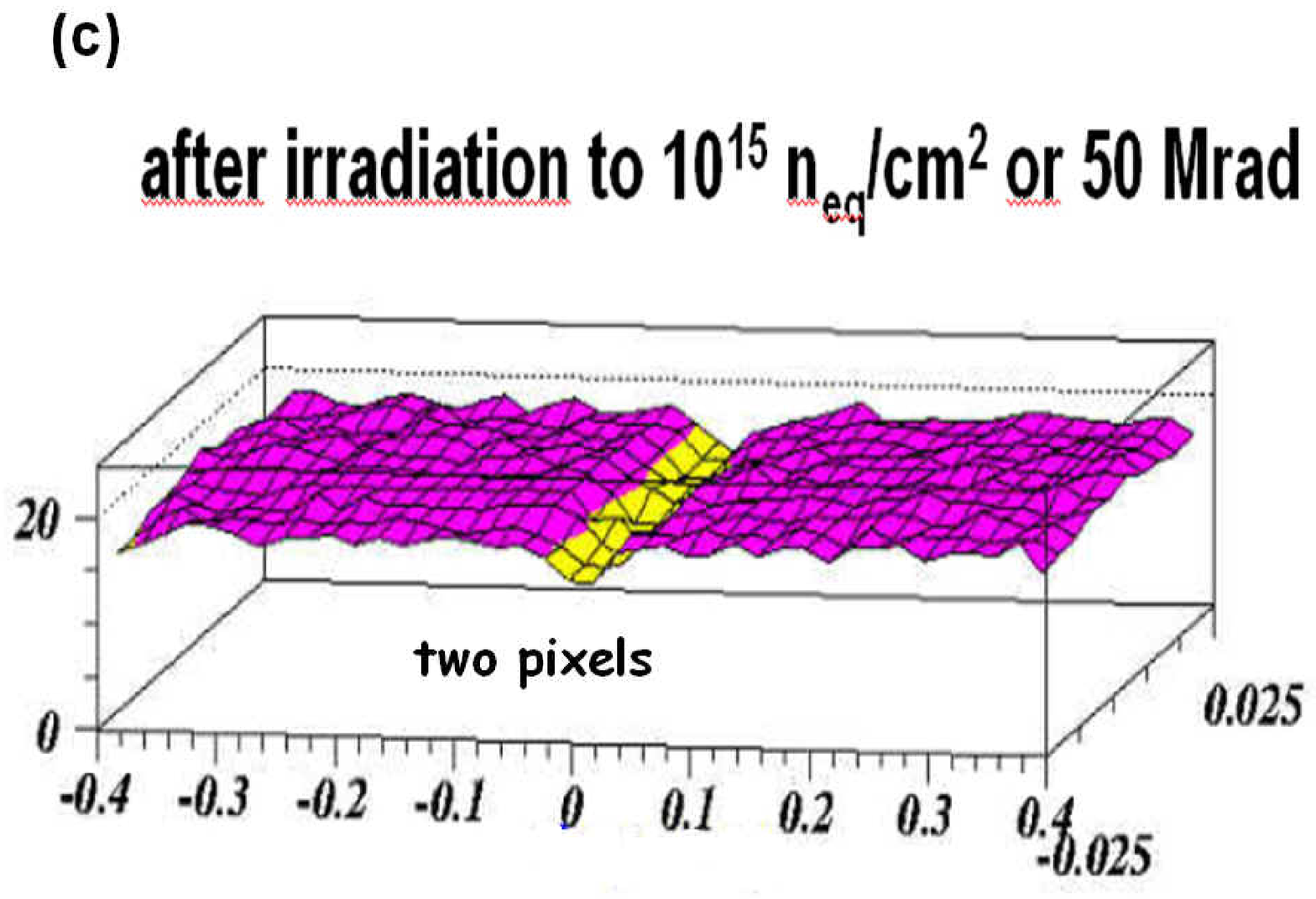}
\end{center}
\caption[]{\label{fig1} (a) Effective doping concentration and
depletion voltage as a function of the fluence, (b) ATLAS silicon
pixel wafer with 3 tiles, and (c) charge collection efficiency in
two adjacent pixels after irradiation.} \label{fig3}
\end{figure}

In the \emph{hybrid pixel technique} sensor and FE-chips are
separate parts of the detector module connected by small
conducting bumps applied by using the bumping and flip-chip
technology. All LHC-collider-detectors \cite{ATLAS,CMS,ALICE}
ALICE, ATLAS, and CMS, LHCb for the RICH system \cite{LHCb}, as
well as some fixed target experiments (NA60\cite{NA60} at CERN and
BTeV\cite{BTEV} at Fermilab) employ the hybrid pixel technique to
build large scale ($\sim$m$^2$) pixel detectors. Pixel area sizes
are typically $50 \mu$m $\times 400 \mu$m as for ATLAS or $100
\mu$m $\times 150 \mu$m as for CMS. The detectors are arranged in
cylindrical barrels of $2-3$ layers and disks covering the forward
and backward regions. The main purpose of the detectors at LHC is
(a) the identification of short lived particles (e.g. b-tagging
for Higgs and SUSY signals), (b) pattern recognition and event
reconstruction and (c) momentum measurement. The detectors need to
withstand a total (10 yr) particle fluence of $10^{15}$n$_{eq}$
corresponding to a radiation dose of about $50$ Mrad. The
discovery that oxygenated silicon is radiation hard with respect
to the non-ionizing energy loss of protons and pions
\cite{oxysilicon} saves pixel detectors at the LHC for which the
radiation is most severe due to their proximity to the interaction
point. $n^+$ electrode in n-bulk material sensors have been chosen
to cope with the fact that type inversion occurs after about
$\Phi_{eq} = 2.5 \times 10^{13}$cm$^{-2}$. After type inversion
the $pn$-diode sits on the electrode side thus allowing the sensor
to be operated partially depleted. Figure \ref{fig1}(a) shows the
development of the effective doping concentration and the
depletion voltage $V_{dep}$ as a function of the fluence expected
for 10 years of operation at the LHC (see also \cite{Krasel03}).
Figures \ref{fig1}(b) and (c) show the ATLAS pixel sensor wafer
with 3 tiles \cite{ATLAS-Sensor} and the detector response after
irradiation over two adjacent pixels showing a very homogeneous
charge collection except for some tolerable loss at the outer
edges and in an area between the pixels. These are due to the bump
contact and a bias grid, respectively, which has been designed to
allow testing of the sensors before bonding them to the
electronics ICs.


%

%

%

The challenge in the design of the front-end pixel electronics
\cite{blanquart03, blanquart-portland} can be summarized by the
following requirements: low power ($<$ 50$\mu$W per pixel), low
noise and threshold dispersion (together $<$ 200e), zero
suppression in every pixel, on-chip hit buffering, and small
time-walk to be able to assign the hits to their respective LHC
bunch crossing. The pixel groups have reached these goals in
several design iterations using first radiation-\emph{soft}
prototypes, then dedicated radhard designs, and finally using deep
submicron technologies \cite{blanquart-portland}. While CMS uses
analog readout of hits up to the counting house, ATLAS obtains
pulse height information by means of measuring the \emph{time over
threshold} (ToT) for every hit.

The chip and sensor connection is done by fine pitch bumping and
subsequent flip-chip which is achieved with either PbSn (solder)
or Indium bumps at a failure rate of $< 10^{-4}$. Figure
\ref{fig3} shows rows of $50 \mu$m pitch bumps obtained by these
techniques.

\begin{figure}[htb]
\includegraphics[width=0.48\textwidth]{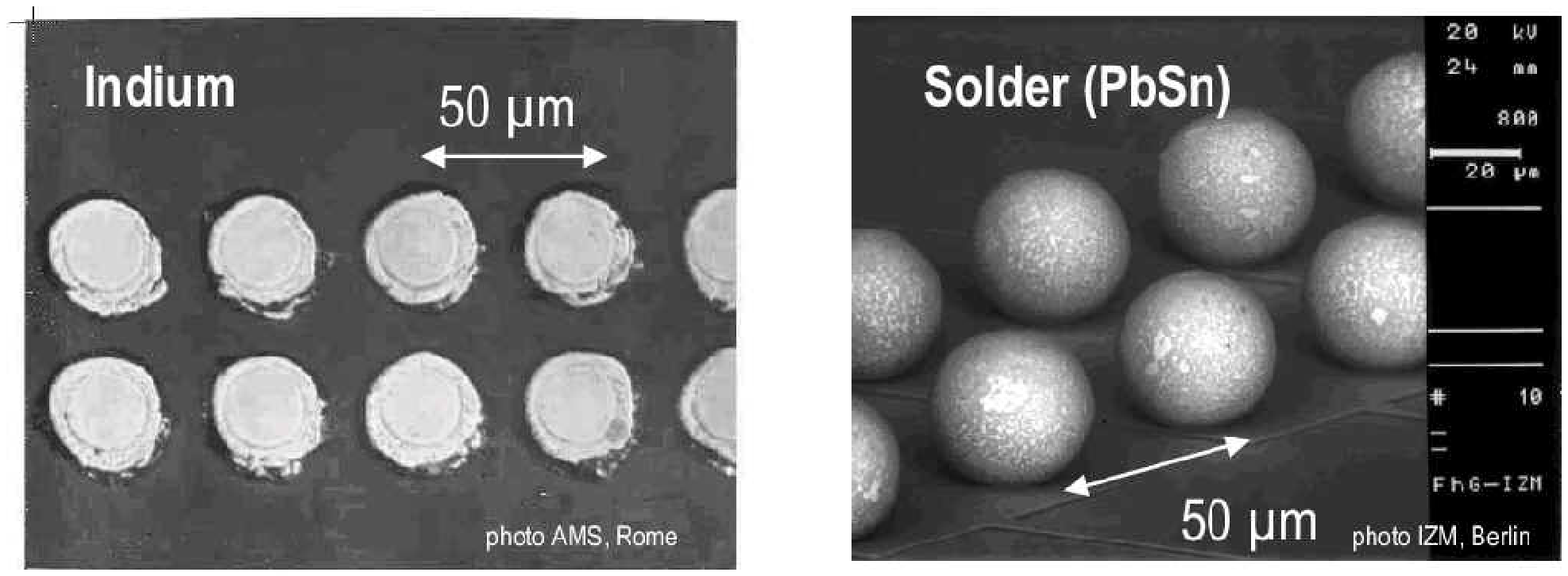}
  \caption{(left) Indium (Photo AMS, Rome) and (right) solder
  (PbSn, Photo IZM, Berlin) bump rows with $50 \mu$m pitch.}
  \label{fig3}
\end{figure}

Indium bumping is done using a wet lift-off technique applied on
both sides (sensor and IC) \cite{In-bumping}. The connection is
obtained by thermo-compression. The Indium joint is comparatively
soft and the gap between IC and sensor is about $6 \mu$m. PbSn
bumps are applied by electroplating \cite{PbSn-bumping}. Here the
bump is galvanically grown on the chip wafer only. The bump is
connected by flip-chipping to an under-bump metallization on the
sensor substrate pixel. Both technologies have been successfully
used with 8" IC-wafers and 4" sensor wafers.

In the case of ATLAS a \emph{module} of 2.1x6.4 cm$^2$ area
consists of FE-chips bump-connected to one silicon sensor. The I/O
lines of the chips are connected via wire bonds to a kapton flex
circuit glued atop the sensor. The flex houses a module control
chip (MCC) responsible for front end time/trigger control and
event building. The total thickness at normal incidence is in
excess of $2 \%$ $X_0$. Figure \ref{fig3} shows a $^{241}$Am (60
keV $\gamma$) source scan of an ATLAS module with 46080 pixels and
an amplitude spectrum obtained using the ToT analog information
without clustering of pixels.

\begin{figure}[htb]
\includegraphics[width=0.48\textwidth]{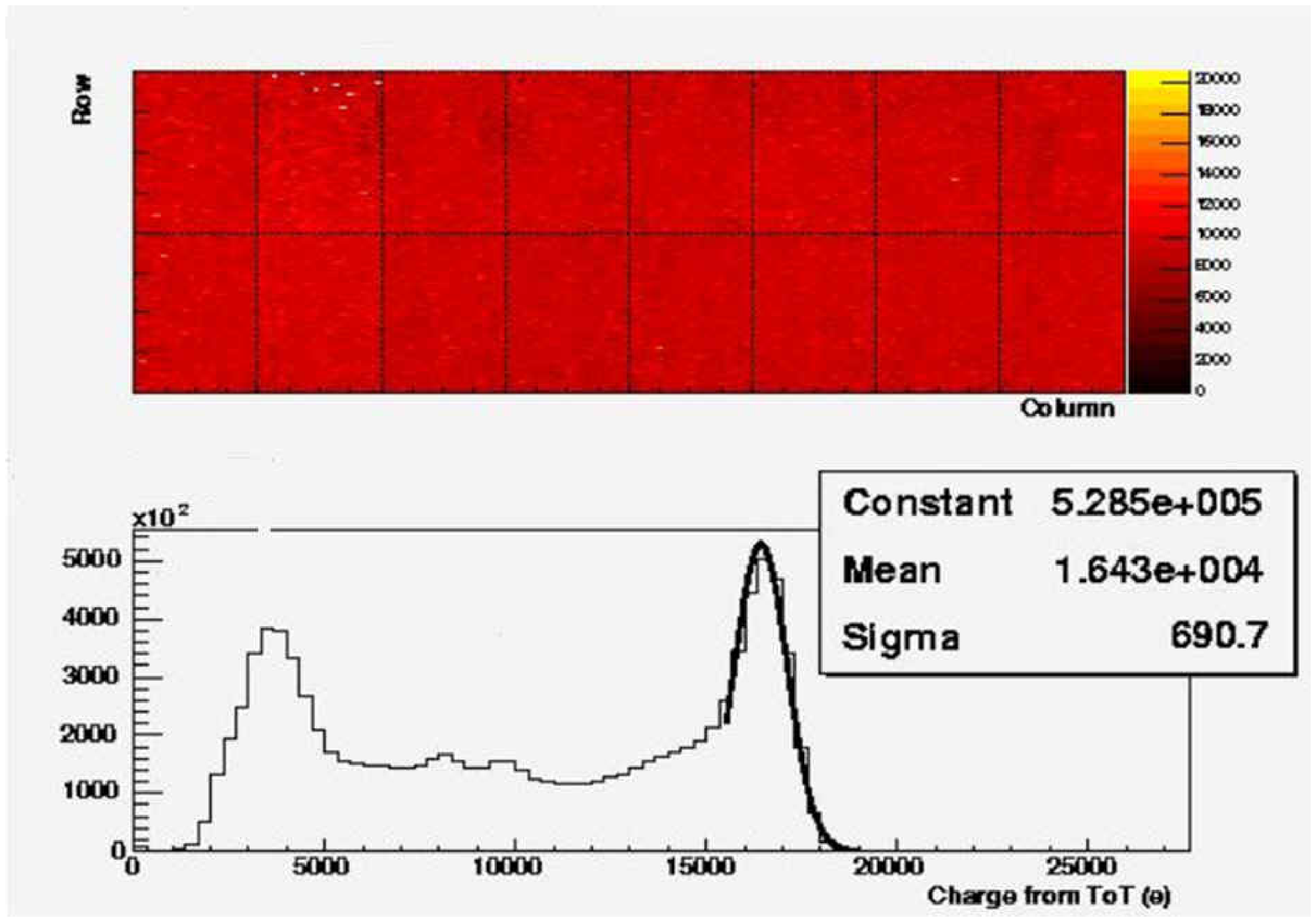}
  \caption{(top) Scan of an ATLAS pixel module (46080 pixels) with a
  $^{241}$Am (60 keV $\gamma$) source and (bottom) the corresponding
  amplitude spectrum obtained using the signal amplitude via
  time-over-threshold (ToT) without clustering.}
  \label{fig3}
\end{figure}

\begin{figure}[htb]
\includegraphics[width=0.48\textwidth]{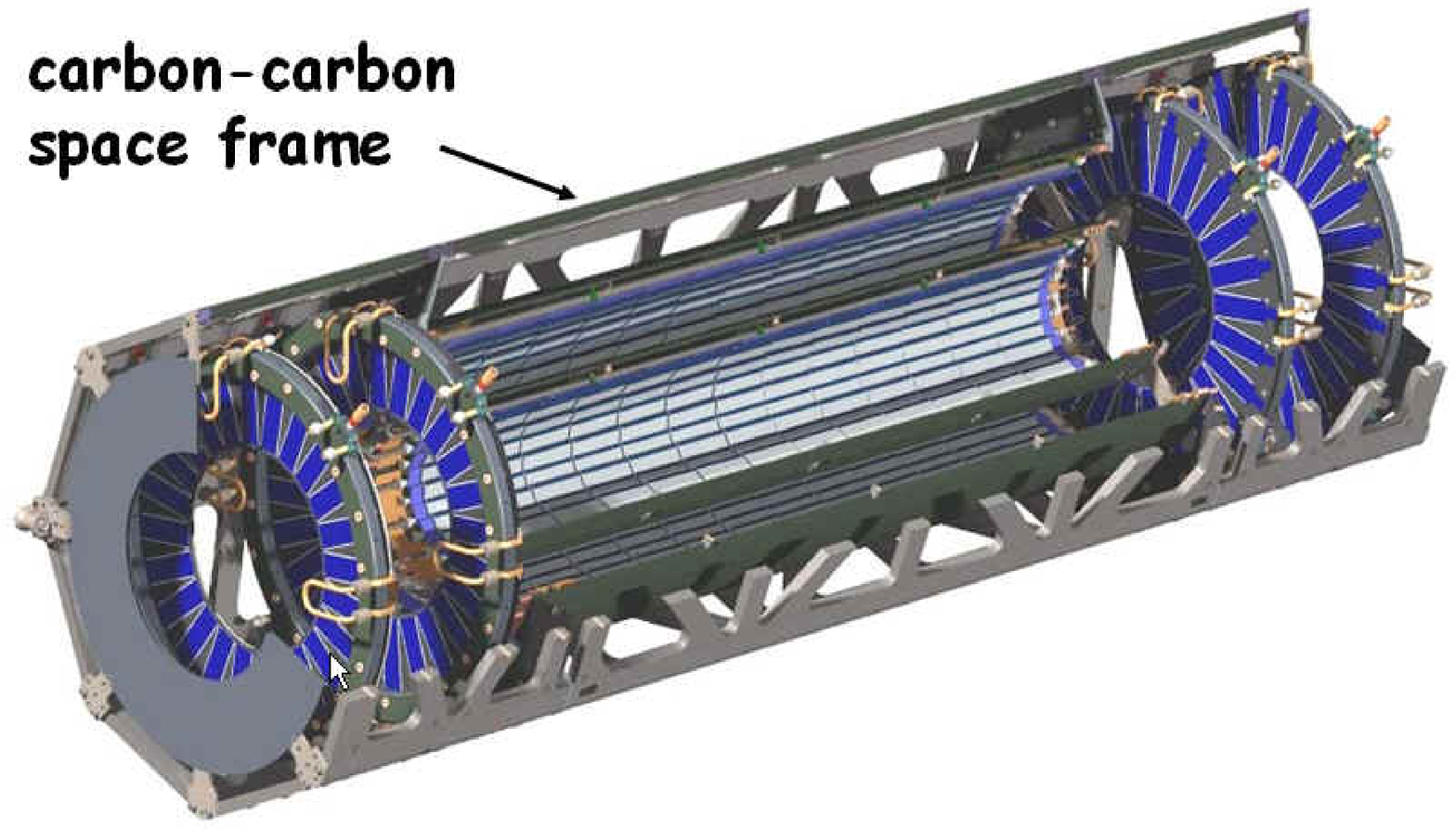}
  \caption{View of the ATLAS pixel detector in its present descoped version with
  two barrel layers and 2x3 disks.}
  \label{fig4}
\end{figure}

The modules are positioned by dedicated robots on carbon-carbon
ladders (staves) and cooled by evaporation of a fluorinert liquid
($C_3F_8$) at an input temperature below $-20 ^{o}C$ in order to
maintain the entire detector below $-6 ^{o}C$ to minimize the
damage induced by radiation. This operation requires pumping and
the cooling tubes must stand $16$ bar pressure if pipe blocking
occurs. All detector components must survive temperature cycles
between $-25 ^{o}C$ and room temperature. The LHC pixel detectors
are presently in production. A sketch of the final ATLAS pixel
detector is shown in fig. \ref{fig4}.

\section{Trends in Hybrid Pixel Detectors}
To improve the performance of Hybrid Pixel detectors in certain
aspects several ideas and developments are being pursued.

\subsection{HAPS}
\begin{figure}[htb]
\includegraphics[width=0.22\textwidth]{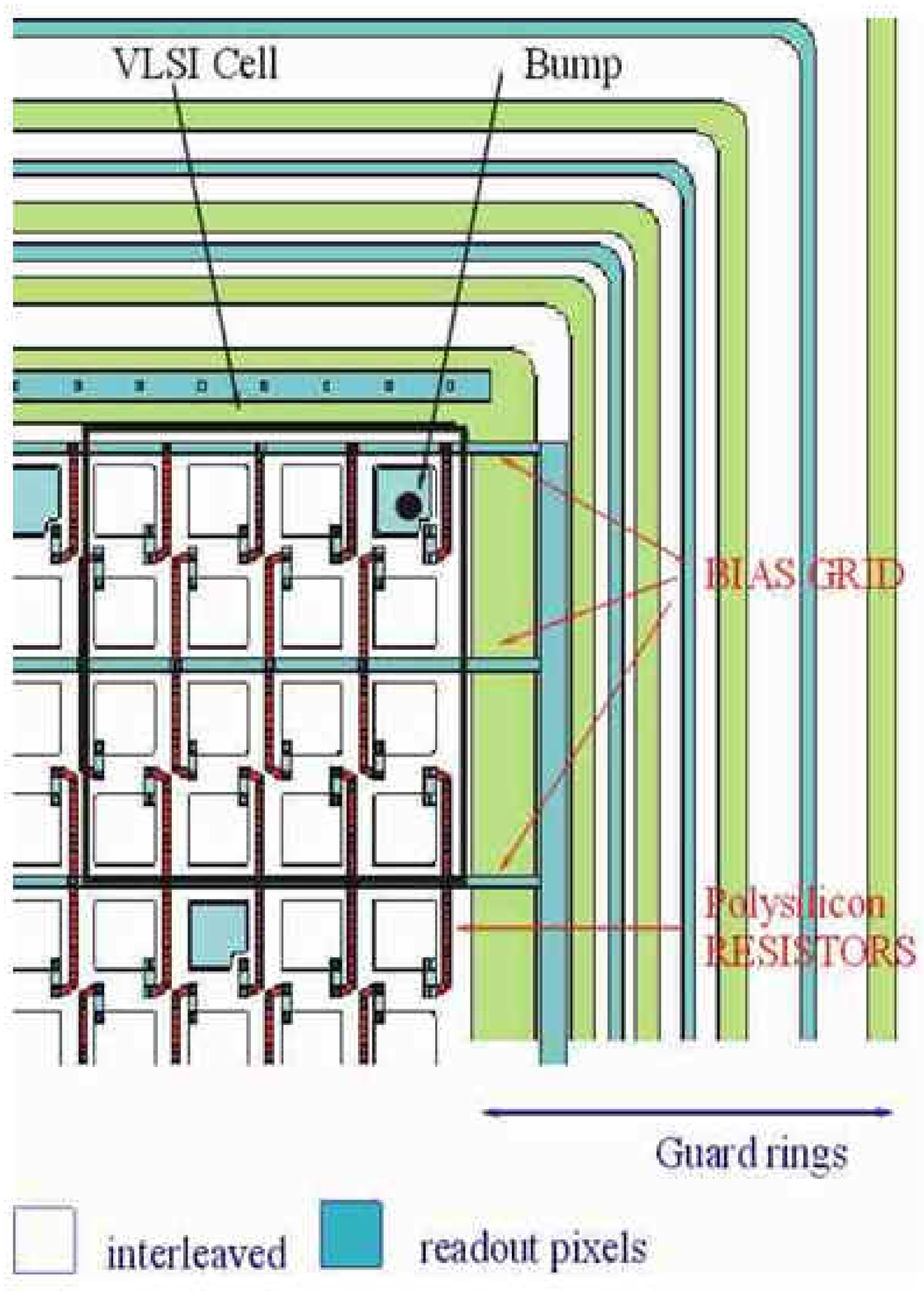}
\includegraphics[width=0.26\textwidth]{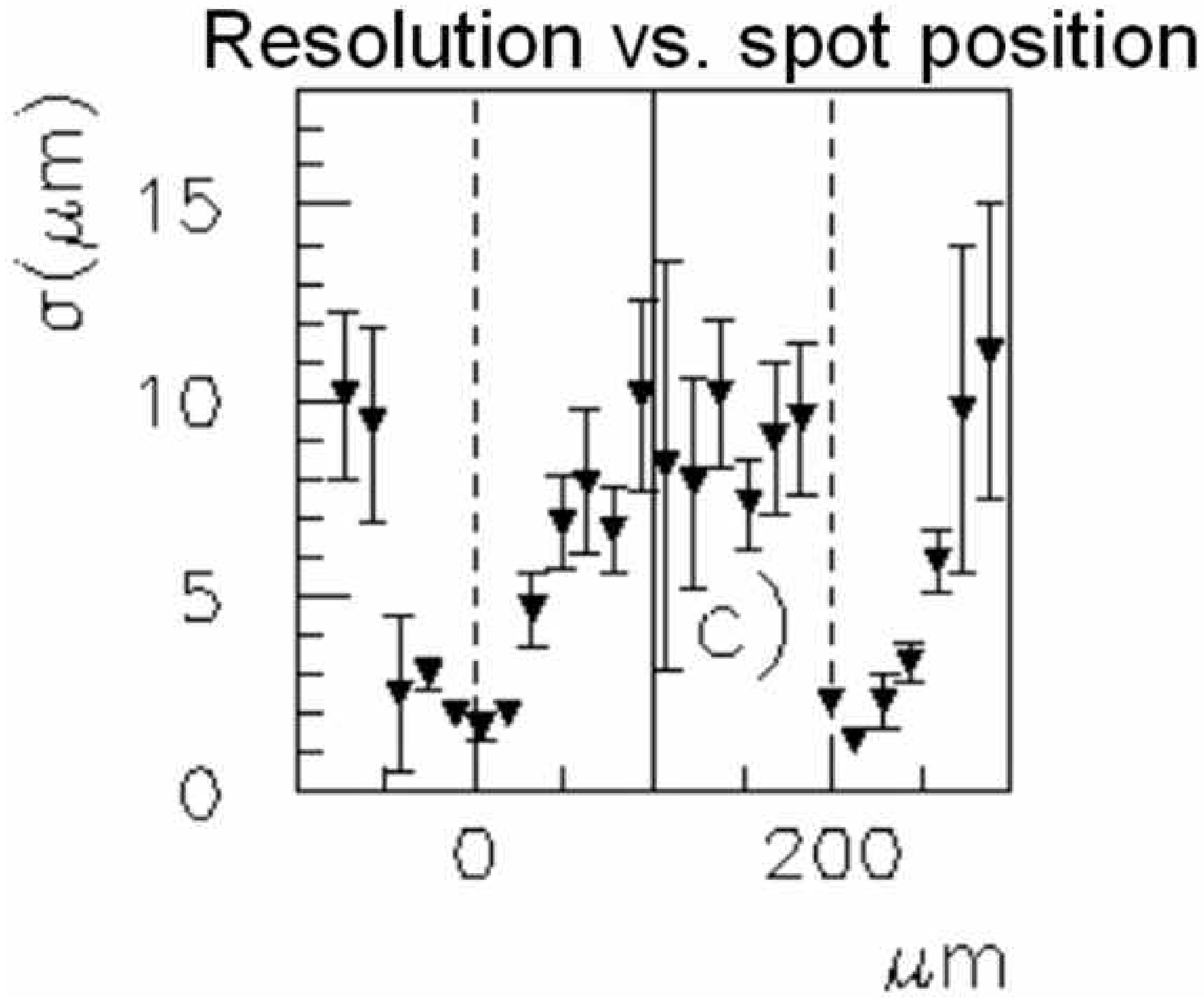}
  \caption{(left) Layout of a HAPS sensor with different pixel and readout pitches, (right)
  measured spatial resolutions by scanning with a fine focus laser.}
  \label{fig5}
\end{figure}

Hybrid (Active) Pixel Sensors (HAPS) \cite{HAPS} exploit
capacitive coupling between pixels to obtain smaller pixel cells
and pixel pitch with a larger readout pitch resulting in
interleaved pixels. The pixel pitch is designed for best spatial
resolution using charge sharing between neighbors while the
readout pitch is tailored to the needs for the size of the
front-end electronics cell. This way resolutions between 3$\mu$m
and 10$\mu$m can be obtained with pixel (readout) pitches of
100$\mu$m (200$\mu$m) as is shown in fig. \ref{fig5}(b) together
with the layout of a prototype sensor in fig. \ref{fig5}(a).

\subsection{MCM-D}
\begin{figure}[htb]
\includegraphics[width=0.48\textwidth]{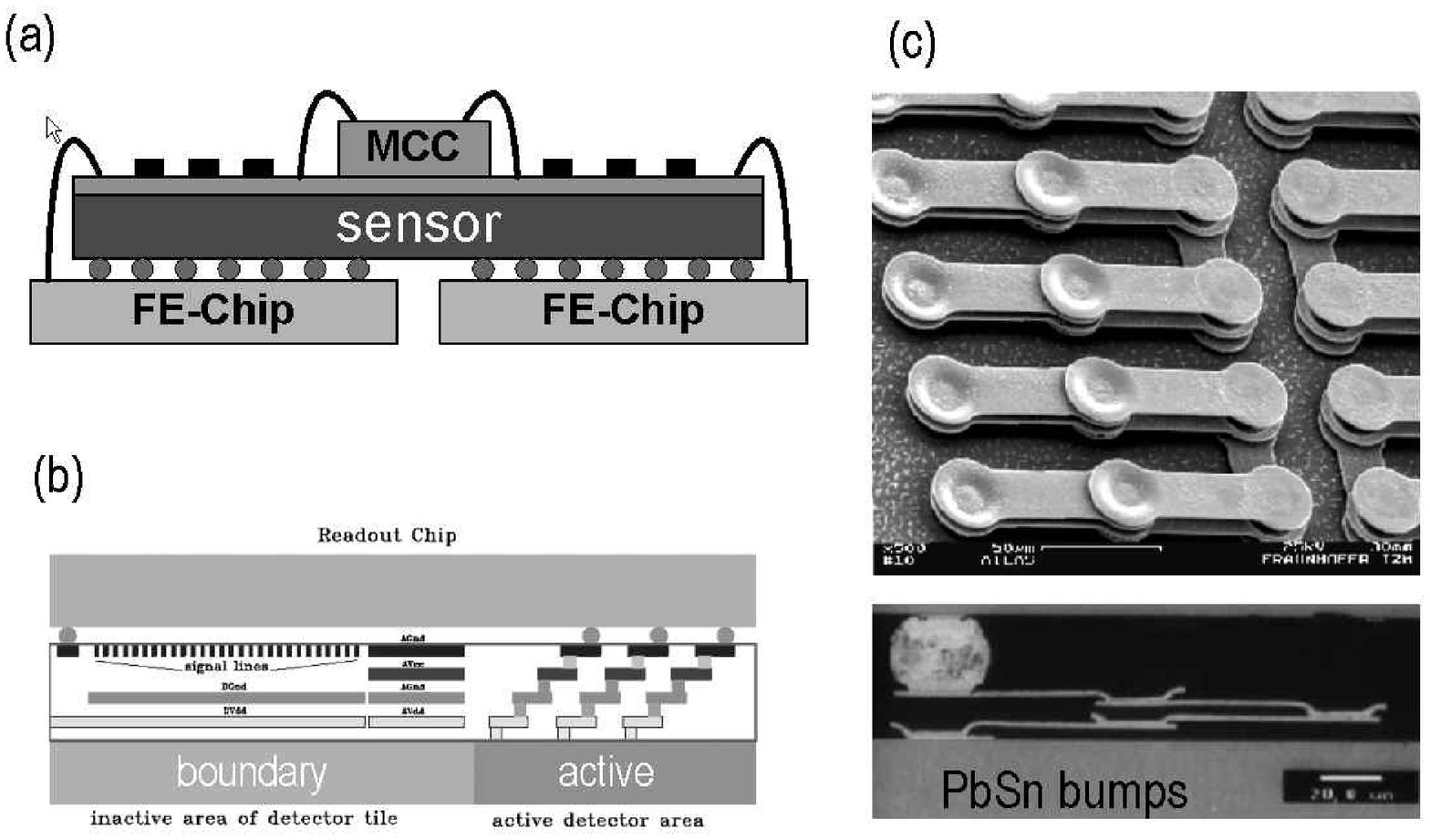}
  \caption{(a) Schematic view of a hybrid pixel module showing ICs
  bonded via bump connection to the sensor, the flex hybrid kapton
  layer atop the sensor with mounted electrical components,
  and the module control chip (MCC). Wire bond connections are needed as indicated.
  (b) Schematic layout of a MCM-D pixel module. (c) SEM
  photograph of a MCM-D via structure (top) and a cross section
  after bump deposition (bottom).
  }
  \label{fig6}
\end{figure}

The present hybrid-pixel modules of the LHC experiments use an
additional flex-kapton fine-print layer on top of the Si-sensor
(fig. \ref{fig6}(a)) to provide power and signal distribution to
and from the module front-end chips. An interesting alternative to
the flex-kapton solution is the so-called Multi-Chip-Module
Technology deposited on Si-substrate (MCM-D) \cite{MCM-D}. A
multi-conductor-layer structure is built up on the silicon sensor.
This allows to bury all bus structures in four layers in the
inactive area of the module thus avoiding the kapton flex layer
and any wire bonding at the expense of a small thickness increase
of $0.1 \% \ X_0$. In addition, the extra freedom in routing makes
it possible to design pixel detectors which have the same pixel
dimensions throughout the sensor, i.e. without larger edge pixels
usually necessary to accommodate the sensor area which remains
uncovered by electronics in between chips. Figure \ref{fig6}(b)
illustrates the principle and fig. \ref{fig6}(c) shows scanning
electron microphotographs (curtesy IZM, Berlin) of a via structure
made in MCM-D technology.

\subsection{Active edge 3-D silicon}
The features of doped Si allow interesting geometries with respect
to field shapes and charge collection. So-called 3D silicon
detectors have been proposed \cite{3D-parker} to overcome several
limitations of conventional planar Si-pixel detectors, in
particular in high radiation environments, in applications with
inhomogeneous irradiation and in applications which require a
large active/inactive area ratio such as protein crystallography
\cite{3Dwestbrook}.
\begin{figure}
\begin{center}
\includegraphics[width=0.30\textwidth]{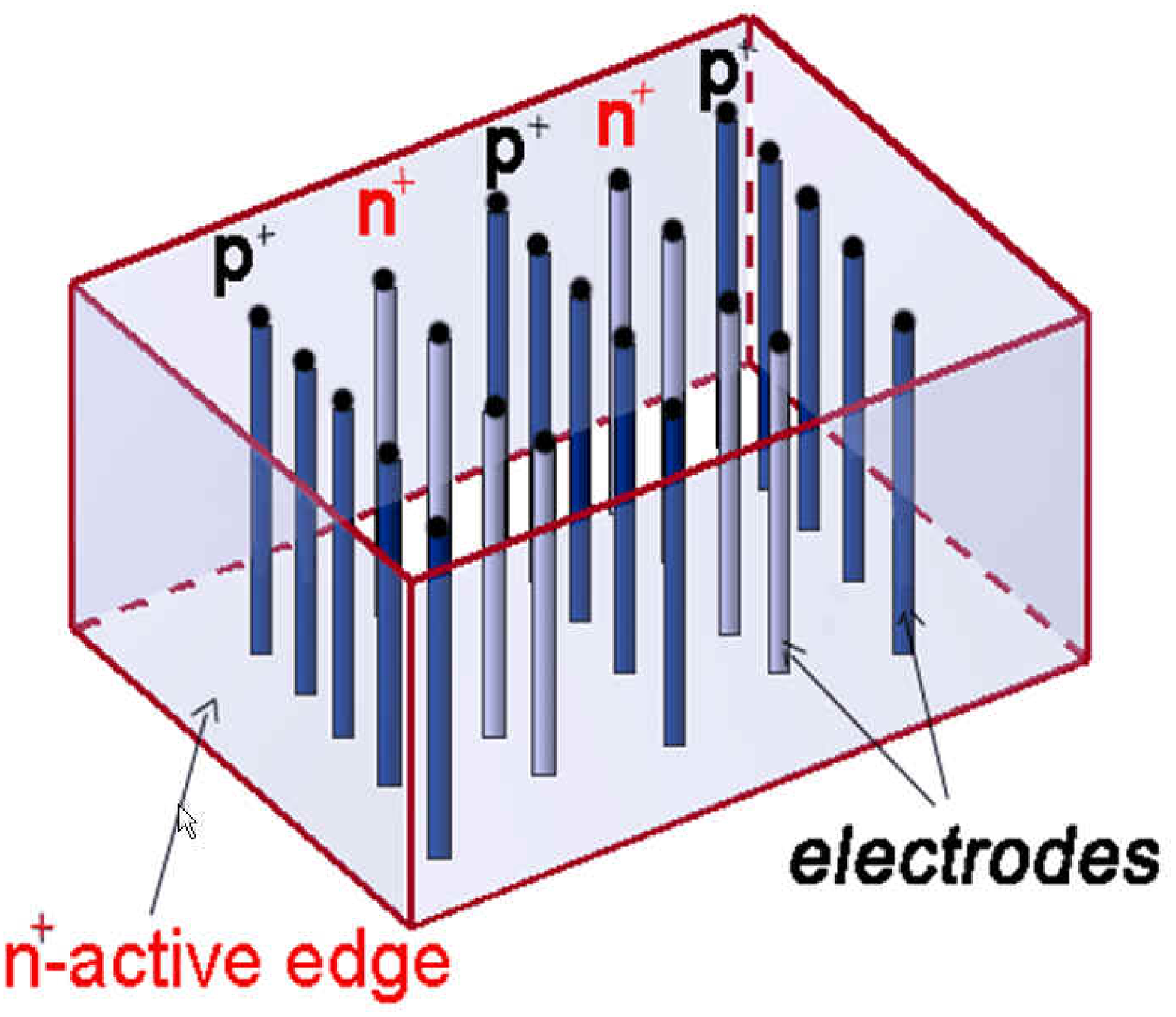}
\vskip 0.5cm
\includegraphics[width=0.48\textwidth]{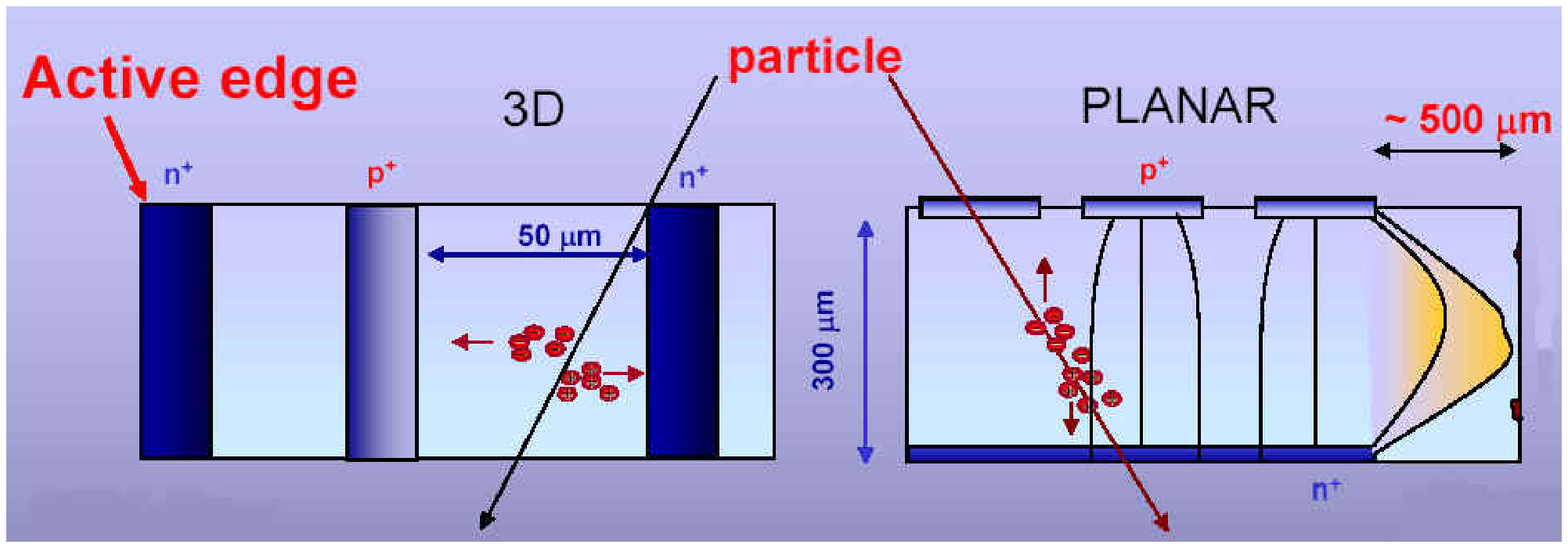}
\end{center}
\caption[]{(top) Schematic view of a 3D silicon detector, (bottom)
comparison of the charge collection in 3D-silicon and conventional
planar electrode detectors.} \label{fig7}
\end{figure}
A a 3D-Si-structure (fig. \ref{fig7}(a)) is obtained by processing
the n$^+$ and p$^+$ electrodes into the detector bulk rather than
by conventional implantation on the surface. This is done by
combining VLSI and MEMS (Micro Electro Mechanical Systems)
technologies. Charge carriers drift inside the bulk parallel to
the detector surface over a short drift distance of typically
50$\mu$m. Another feature is the fact that the edge of the sensor
can be a collection electrode itself thus extending the active
area of the sensor to within about 10$\mu$m to the edge. This
results from the undepleted depth at this electrode. Edge
electrodes also avoid inhomogeneous fields and surface leakage
currents which usually occur due to chips and cracks at the sensor
edges. This is seen in fig. \ref{fig7}(b) which shows the two
structures in comparison. The main advantages of 3D-silicon
detectors come from the fact that the electrode distance is short
(50$\mu$m) in comparison to conventional planar devices at the
same total charge, resulting in a fast (1-2 ns) collection time,
low ($<$ 10V) depletion voltage and, in addition, a large
active/inactive area ratio of the device.

The technical fabrication is much more involved than for planar
processes and requires a bonded support wafer and reactive ion
etching of the electrodes into the bulk. A compromise between 3D
and planar detectors, so called planar-3D detectors maintaining
the large active area, use planar technology but with edge
electrodes \cite{3Dplanar}, obtained by diffusing the dopant from
the deeply etched edge and then filling it with poly-silicon.
Prototype detectors using strip or pixel electronics have been
fabricated and show encouraging results with respect to speed (3.5
ns rise time) and radiation hardness ($>$ 10$^{15}$
protons/cm$^2$) \cite{3DPortland}.

\subsection{Diamond Pixel Sensors}
Diamond as a potentially very radiation hard material has been
explored intensively by the R$\&$D collaboration RD42 at CERN
\cite{RD42}. Charge collection distances up to 300$\mu$m have been
achieved. Diamond pixel detectors have been built with pixel
electronics developed for the ATLAS pixel detector and operated as
single-chip modules in testbeams \cite{diamond_keil}. A spatial
resolution of $\sigma$ = 22$\mu$m has been measured with 50$\mu$m
pixel pitch, compared to about 12$\mu$m with Si pixel detectors of
the same geometry and using the same electronics. This reveals the
characteristic systematic spatial shifts observed in CVD diamond
due to the crystal growth. Grain structures are formed in which
trapped charges at the boundaries produce polarization fields
inside the sensor which cause the observed spatial shifts of the
order of 100-150 $\mu$m.

\section{Hybrid Pixels in Imaging Applications}
\subsection{Counting Pixel Detectors for (Auto-)Radiography}
A potentially new route in radiography techniques is the counting
of individual X-ray photons in every pixel cell. This approach
offers many features which are very attractive for X-ray imaging:
full linearity in the response function and a principally infinite
dynamic range, optimal exposure times and a good image contrast
compared to conventional film-foil based radiography. It thus
avoids over- and underexposed images. Counting pixel detectors
must be considered as a very direct spin-off of the detector
development for particle physics into biomedical applications. The
analog part of the pixel electronics is in parts close to
identical to the one for LHC pixel detectors while the periphery
has been replaced by the counting circuitry \cite{PeFicounter}.

The challenges to be addressed in order to be competitive with
integrating radiography systems are: high speed ($>$ 1 MHz)
counting with a range of at least 15 bit, operation with very
little dead time, low noise and particularly low threshold
operation with small threshold dispersion values. In particular
the last item is important in order to allow homogeneous imaging
of soft X-rays. It is also mandatory for a differential energy
measurement, realized so far as a double threshold with energy
windowing logic \cite{MPEC-windowing,MEDIPIX2}. A differential
measurement of the energy, exploiting the different shapes of
X-ray spectra behind for example tissue or bone, can enhance the
contrast performance of an image. Finally, for radiography high
photon absorption efficiency is mandatory, which renders the not
easy task of high Z sensors and their hybridization necessary.

\begin{figure}
\begin{center}
\includegraphics[width=0.48\textwidth]{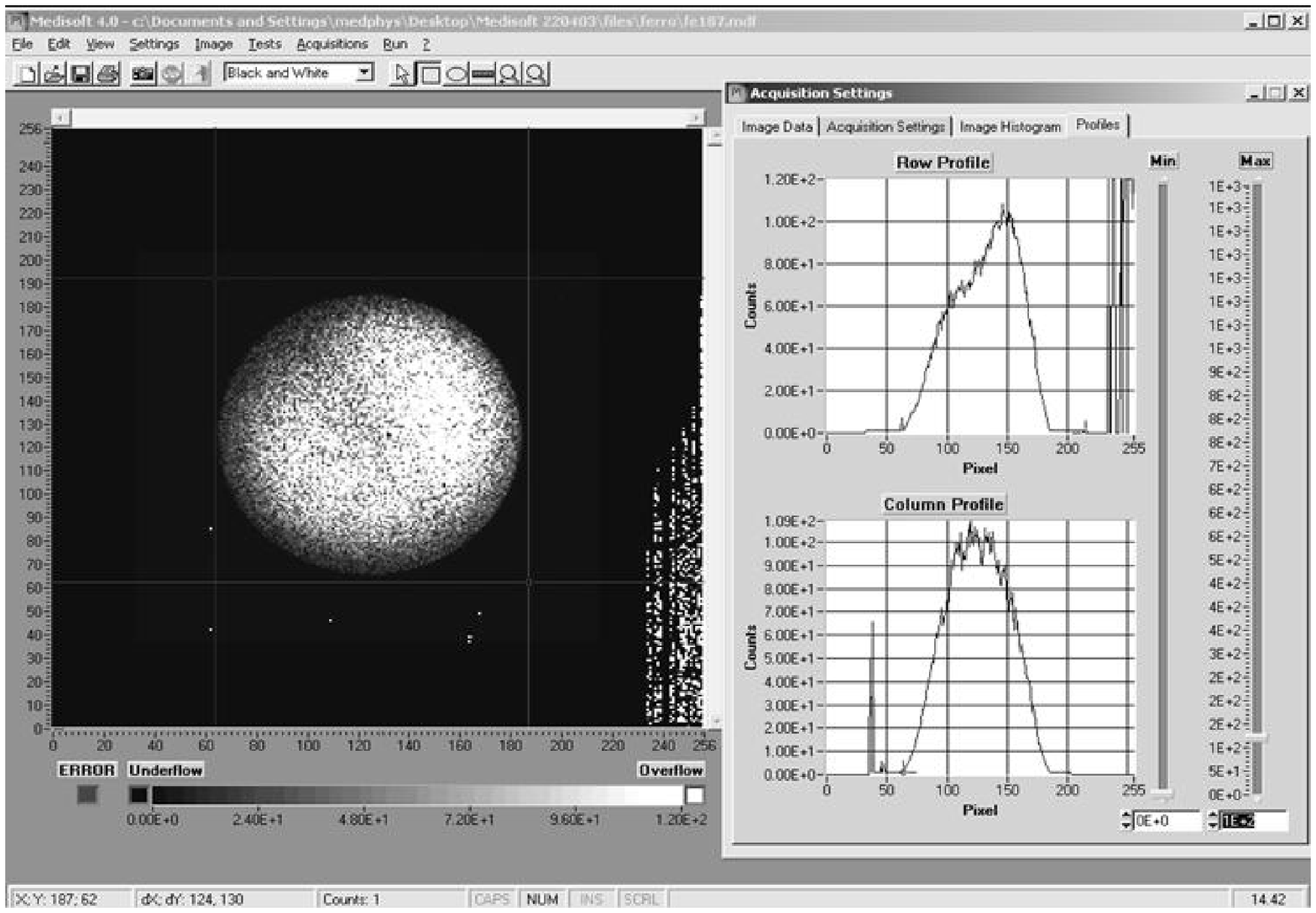}
\end{center}
\caption[]{Image of a $^{55}$Fe point source taken with the
MEDIPIX2 counting chip and a silicon sensor (14x14
mm$^2$)\cite{MEDIPIX_Portland}.} \label{fig8}
\end{figure}

Progress with counting pixel systems have been reported at this
conference \cite{MEDIPIX_Portland,MPEC_Portland,Edling_Portland}.
The MEDIPIX chip developed by the MEDIPIX collaboration
\cite{MEDIPIX,MEDIPIX2} uses $256 \times 256$,
55$\times$55$\mu$m$^2$ pixels fabricated in $0.25 \mu$m
technology, energy windowing via two tunable discriminator
thresholds, and a 13 bit counter. The maximum count rate per pixel
is about 1 MHz. Figure \ref{fig8} shows an image of a $^{55}$Fe
point source obtained with the Medipix2 chip bonded to a 14x14
mm$^2$, 300$\mu$m thick Si sensor \cite{MEDIPIX_Portland}. A
similar image with a single chip CdTe detector and a higher energy
$^{109}$Cd source (122 keV $\gamma$) has also been taken and the
capability to detect low energy beta decays from tritium has been
demonstrated \cite{MEDIPIX_Portland}.

\begin{figure}
\begin{center}
\includegraphics[width=0.25\textwidth]{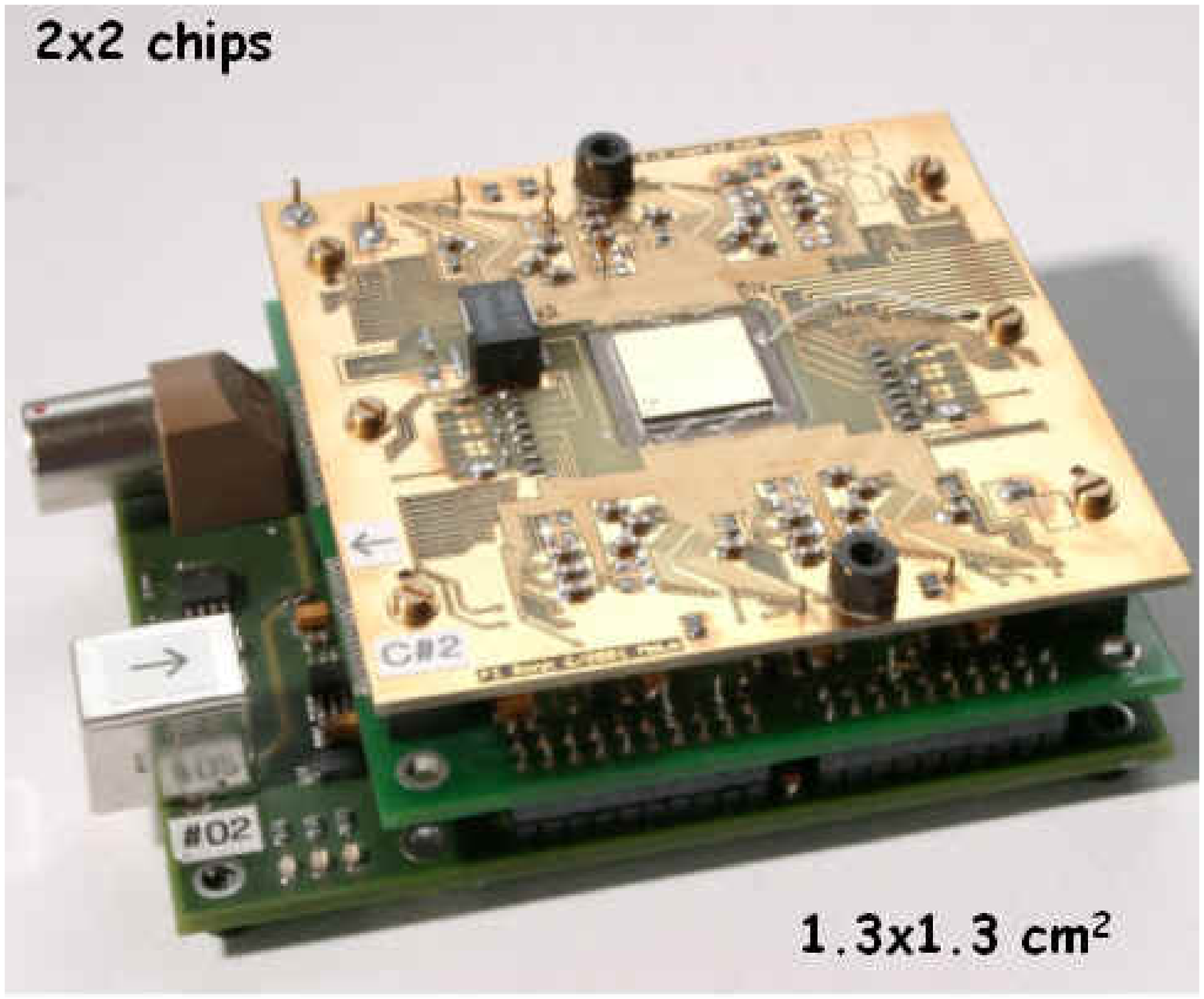}
\includegraphics[width=0.23\textwidth]{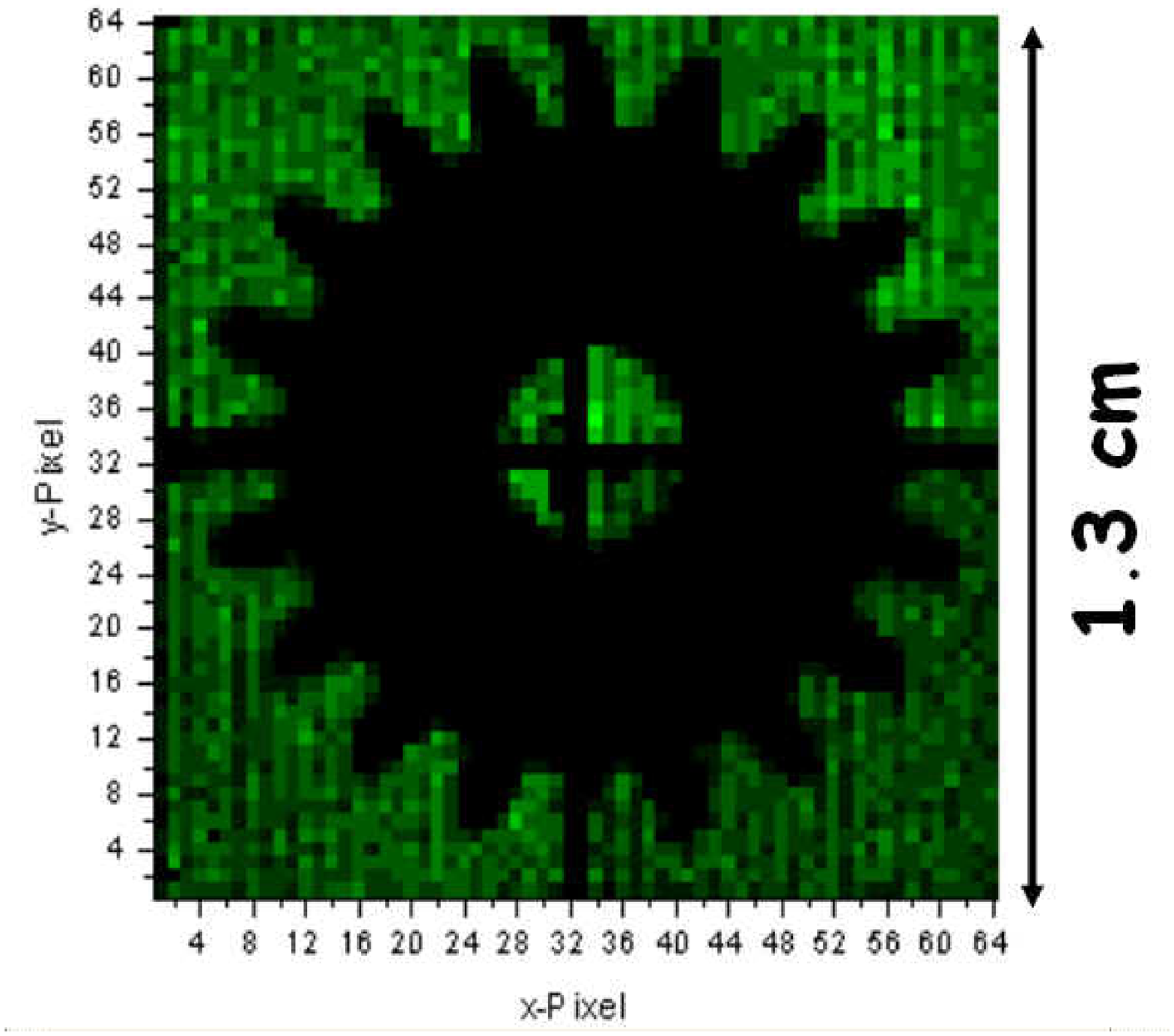}
\end{center}
\caption[]{(left) 2x2 Multi-Chip CdTe Module with counting pixel
electronics using the MPEC chip on a USB based readout board
\cite{MPEC_Portland}, (right) image of a cogwheel taken with 60
keV X-rays.} \label{fig9}
\end{figure}

A Multi-Chip module with 2x2 chips using high-Z CdTe sensors with
the MPEC chip \cite{MPEC_Portland} is shown in fig. \ref{fig9}.
The MPEC chip features $32 \times 32$ pixels
(200$\times$200$\mu$m$^2$), double threshold operation, 18-bit
counting at $\sim$1 MHz per pixel as well as low noise values
($\sim$120e with CdTe sensor) and threshold dispersion ($21$e
after tuning) \cite{MPEC-refs,MPEC_Portland}. A technical issue
here is the bumping of individual die CdTe sensors which has been
solved using Au-stud bumping with In-filling \cite{MPEC-CdTe}.

A challenge for counting pixel detectors in radiology is to build
large area detectors. Commercially available integrating
pixellated systems such as flat panel imagers
\cite{flat-panel,a-Se} constitute the levelling scope.

\subsection{Protein Crystallography}
\begin{figure}
\begin{center}
\includegraphics[width=0.45\textwidth]{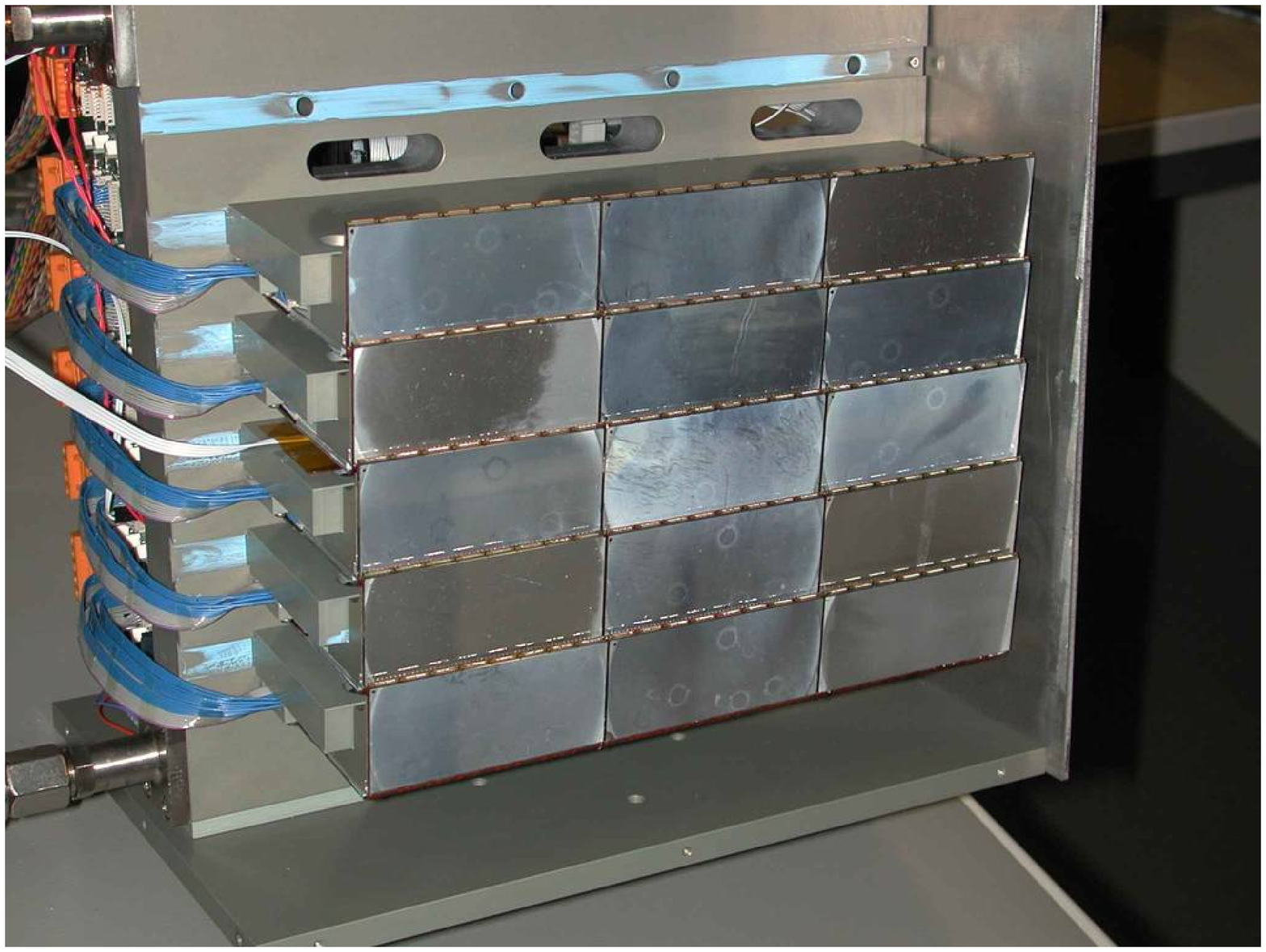}
\vskip 0.5cm
\includegraphics[width=0.45\textwidth]{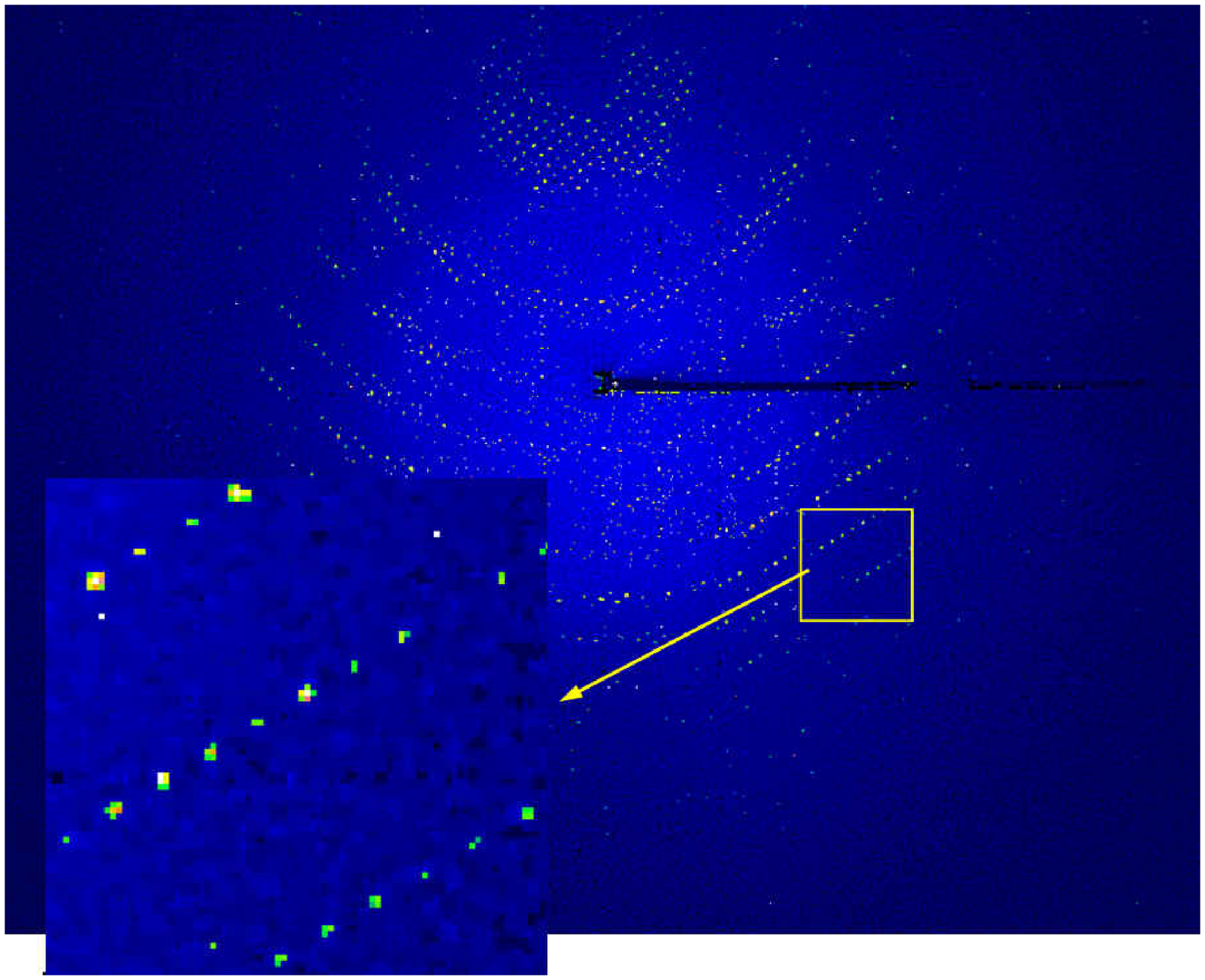}
\end{center}
\caption[]{(top) Photograph of the 20x24 cm$^2$ large PILATUS 1M
detector for protein crystallography using counting hybrid pixel
detector modules, (bottom) image of Lysozyme taken with PILATUS
\cite{PILATUS_Portland}, the insert shows that Bragg spots are
often contained in one pixel.} \label{fig10}
\end{figure}

Hybrid Pixel detectors as counting imagers lead the way to new
classes of experiments in protein crystallography with synchrotron
radiation \cite{Graafsma_Portland}. Here the challenge is to
image, with high rate ($\sim$1-1.5 MHz/pixel) and high dynamic
range, many thousands of Bragg spots from X-ray photons of
typically 12 keV or higher (corresponding to resolutions at the
1$\AA$ range), scattered off protein crystals. The typical spot
size of a diffraction maximum is $100-200 \mu$m, calling for pixel
sizes in the order of $100-300 \mu$m. The high linearity of the
hit counting method and the absence of so-called "blooming
effects", i.e. the response of non-hit pixels in the close
neighborhood of a Bragg maximum, makes counting pixel detectors
very appealing for protein crystallography experiments. A
systematic limitation and difficulty is the problem that
homogeneous hit/count responses in all pixels, also for hits at
the pixel boundaries or between pixels where charge sharing plays
a role must be maintained by delicate threshold tuning. Counting
pixel developments are made for ESRF (Grenoble, France)
\cite{XPAD} and SLS (Swiss Light Source at the Paul-Scherrer
Institute, Switzerland) beam lines. The PILATUS 1M detector
\cite{PILATUS} at the SLS ($217 \mu$m $\times 217 \mu$m pixels) is
made of fifteen 16-chip-modules each covering $8 \times 3.5$
cm$^2$, i.e. a total area of $40 \times 40$cm$^2$. Figure
\ref{fig10}(a) shows a photograph of this detector with 15 modules
and close to 10$^6$ pixels covering an area of 20$\times$24 cm$^2$
\cite{PILATUS_Portland}. It is the first large scale hybrid pixel
detector in operation. Figure \ref{fig10}(b) shows a Bragg image
of a Lysozyme with 10s exposure to 12 keV sychrotron X-rays
\cite{PILATUS_Portland}. Many spots are contained in one pixel
which is the limit of the achievable point spread resolution.

In summary hybrid pixel detectors are {\underline the} present
state of the art in pixel detector technology. For tracking as
well as for imaging applications large ($\sim$m$^2$) detectors are
being built or already in operation (PILATUS). On the negative
side are the comparatively large material budget, the complicated
assembling (hybridization) with potential yield pitfalls, and the
difficulty to obtain high assembly yields with non-wafer scale
high-Z sensor materials. As trends within this technology,
interleaved pixels and MCM-D promise better resolution and more
homogeneous modules, while 3D-silicon detectors can address
applications in high radiation environments or those where a large
active/inactive area ratio is mandatory.

\section{Monolithic and Semi-Monolithic Pixel Detectors}
Monolithic pixel detectors, in which amplifying and logic
circuitry as well as the radiation detecting sensor are one
entity, produced in a commercially available technology, are the
dream of semiconductor detector developers. This will not be
possible in the near future without compromising. Developments and
trends in this direction have much been influenced by R$\&$D for
vertex tracking detectors at future colliders such as a  Linear
$e^+e^-$ Collider \cite{TESLA-TDR}. Very low ($\ll$1$\%$ X$_0$)
material per detector layer, small pixel sizes
($\sim$20$\mu$m$\times 20 \mu$m) and a good rate capability (80
hits/mm$^2$/ms) is required, due to the very intense beamstrahlung
of narrowly focussed electron beams close to the interaction
region which produce electron positron pairs in vast numbers. High
readout speeds of $50$ MHz line rate with 40$\mu$s frame time are
necessary.

\begin{figure}[h]
\begin{center}
\includegraphics[width=0.22\textwidth]{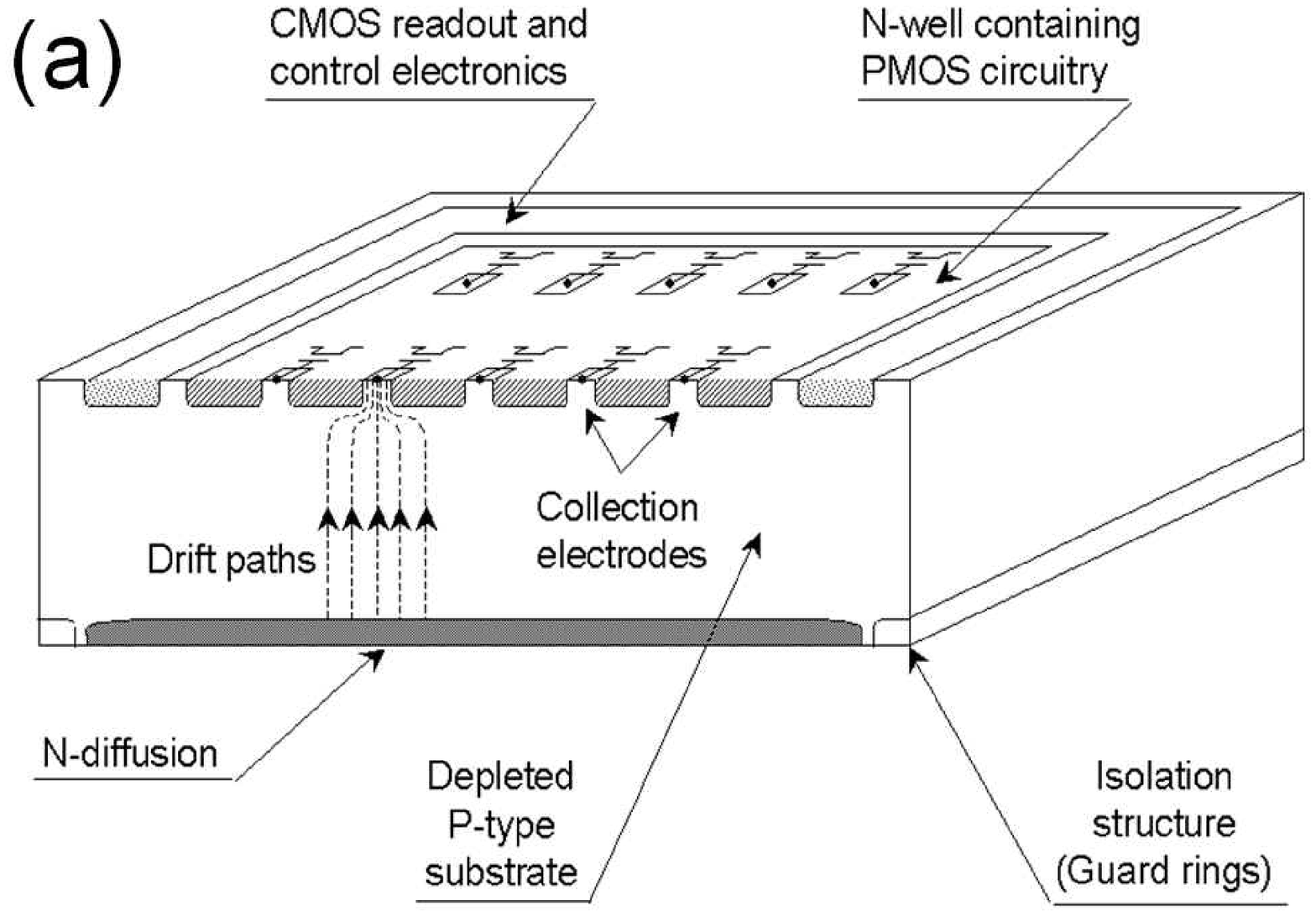}
\includegraphics[width=0.22\textwidth]{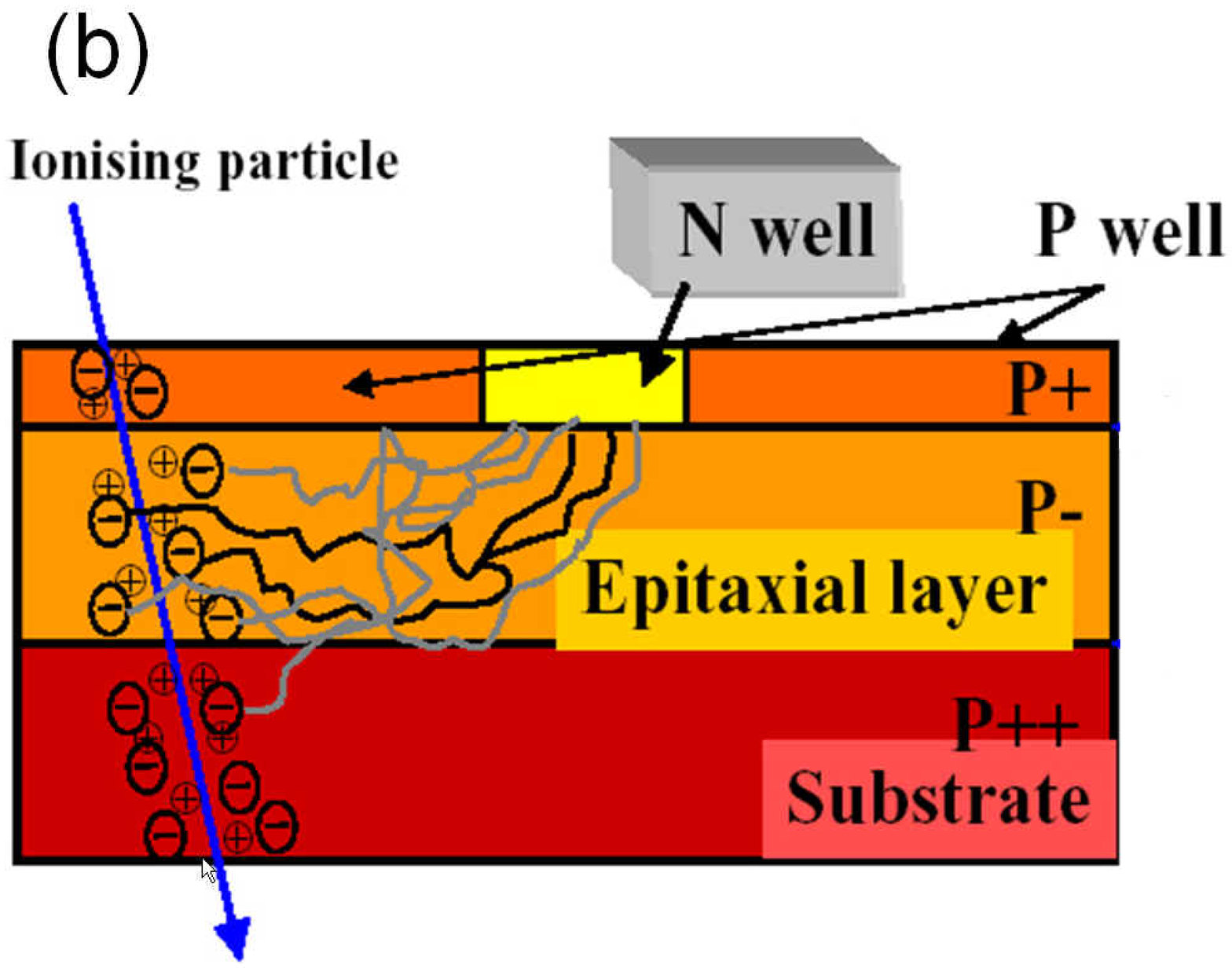}
\end{center}
\caption[]{(a) Sketch of a monolithic pixel detector using a high
resistivity bulk with pMOS transistors at the readout (after
\cite{parker92}), (b) principle of an Monolithic Active Pixel
Sensor (MAPS) \cite{MAPS} targeting CMOS electronics with low
resistivity bulk material. The charge is generated and collected
by diffusion in the few $\mu$m thin epitaxial Si-layer.}
\label{fig11}
\end{figure}

Among the developments of monolithic or semi-monolithic pixel
devices perhaps the following classifying list can be made.
\begin{itemize}
\item \emph{Non-standard CMOS on high resistivity bulk} \\
The first monolithic pixel detector, successfully operated in a
particle beam already in 1992 \cite{parker92}, used a high
resistivity p-type bulk p-i-n detector in which the junction had
been created by an n-type diffusion layer. On one side, an array
of ohmic contacts to the substrate served as collection electrodes
(fig. \ref{fig11}(a)). Due to this only simple pMOS transistor
circuits sitting in n-wells were possible in the active area. The
technology was certainly non-standard and non-commercial. No
further development emerged.
\hfill \\
\item \emph{CMOS technology with charge collection in epi-layer} \\
Commercial CMOS technologies use low resistivity silicon which is
not suited for charge collection. However, in some technologies an
epitaxial silicon layer of a few to 15$\mu$m thickness can be used
\cite{meynants98,MAPS,LEPSI}. The generated charge is kept in the
epi-layer by potential wells at the boundary and reaches an n-well
collection diode by thermal diffusion (fig. \ref{fig11}(b)). The
signal charge is therefore very small ($<$1000e) and low noise
electronics is a real challenge. As this development uses standard
processing technologies it is potentially very cheap. Despite
using CMOS technology the potential of full CMOS circuitry in the
active area is not available (only nMOS) because of the
n-well/p-epi collecting diode which does not permit other n-wells.
\begin{figure}[h]
\begin{center}
\includegraphics[width=0.27\textwidth]{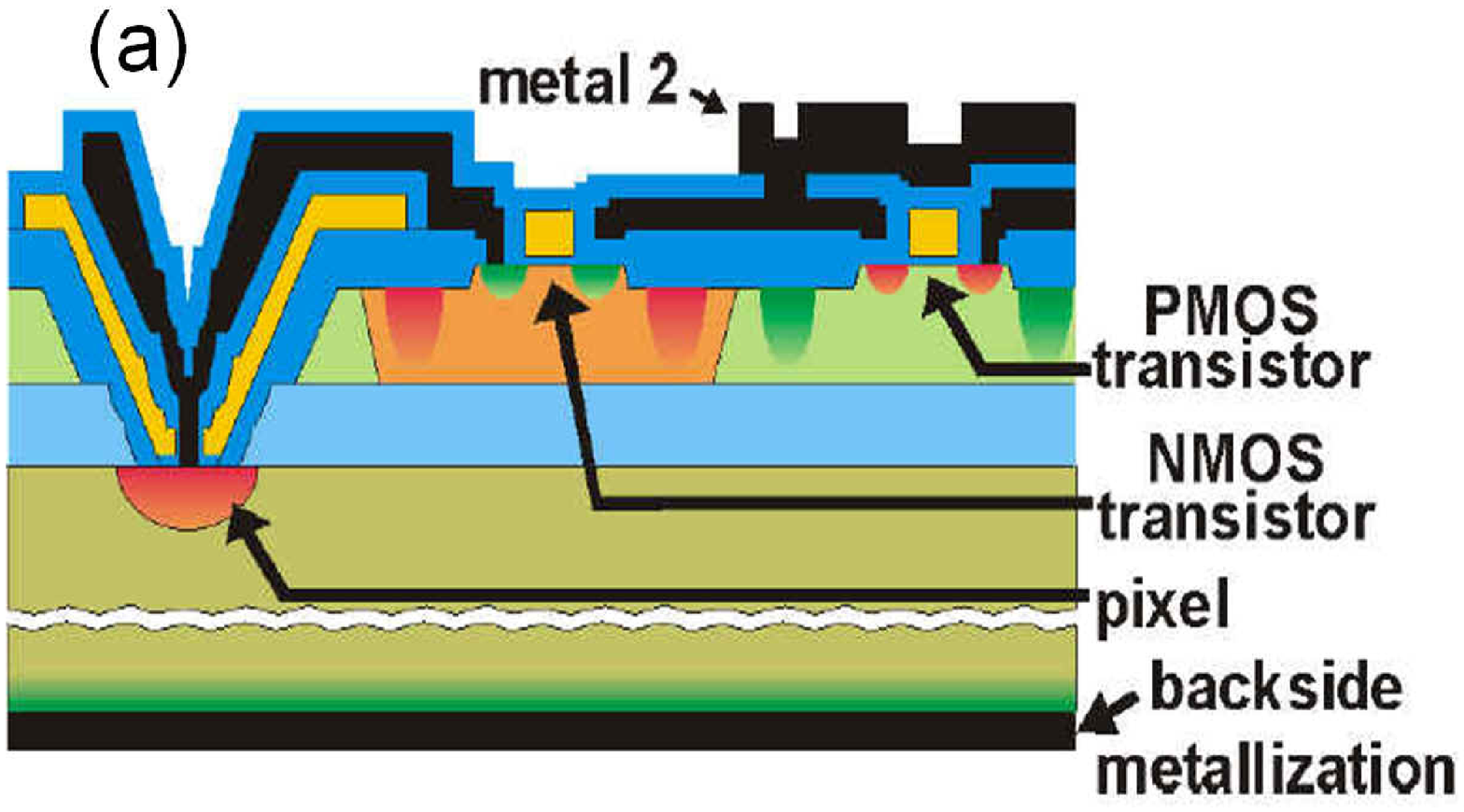}
\includegraphics[width=0.18\textwidth]{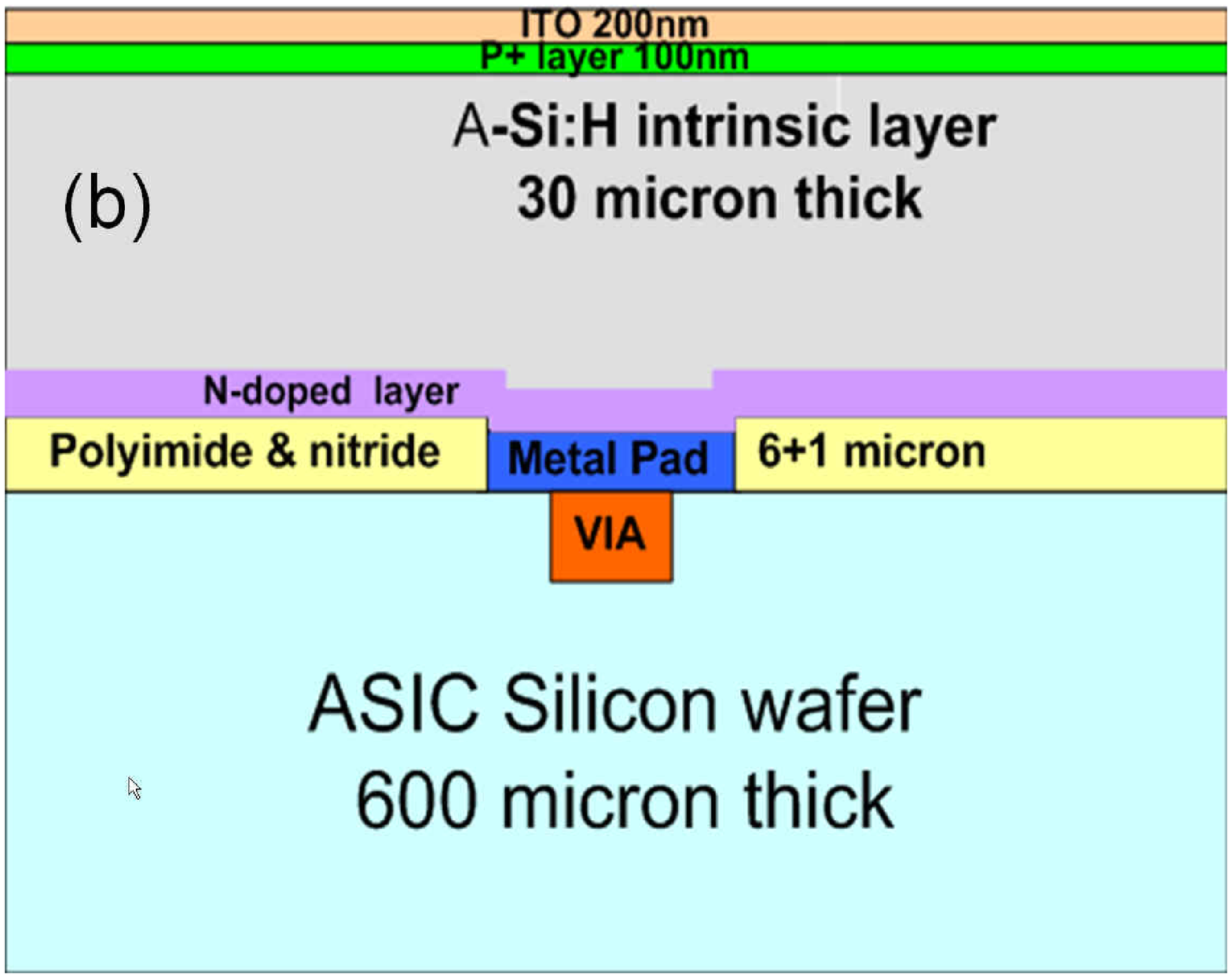}
\end{center}
\caption[]{(a) Cross section through a monolithic CMOS on SOI
pixel detector using high resistivity silicon bulk insolated from
the low resistivity CMOS layer with connecting vias in between
\cite{HAPS-SOI,SOI-Portland}, (b) cross section through a
structure using amorphous silicon on top of standard CMOS VLSI
electronics \cite{theil2001,jarron02}.} \label{fig12}
\end{figure}
\hfill \\
\item \emph{CMOS on SOI (non-standard)} \\
In order to exploit high resistivity bulk material, the authors of
\cite{SOI-Portland} develop pixel sensors with full CMOS circuitry
using the Silicon-on-Insulator (SOI) technology. The connection to
the charge collecting bulk is done by vias (fig. \ref{fig12}(a)).
The technology offers full charge collection in 200--300$\mu$m
high resistive silicon with full CMOS electronics on top. At
present the technology is however definitely not a commercial
standard and the development is still in its beginnings.
\hfill \\
\item \emph{Amorphous silicon on standard CMOS ASICs} \\
Hydrogenated amorphous-Silicon (a-Si:H), where the H-content is up
to 20$\%$, can be put as a film on top of CMOS ASIC electronics.
a-Si:H has been studied as a sensor material long ago and has
gained interest again \cite{theil2001,jarron02} with the
advancement in low noise, low power electronics. The signal charge
collected in the $<$30$\mu$m thick film is in the range of
500-1500 electrons. A cross section through a typical a-Si:H
device is shown in fig. \ref{fig12}(b). From a puristic view it is
more a hybrid technology, but the main disadvantage of the
hybridization connection is absent. The radiation hardness of
these detectors appears to be very high $>$10$^{15}$cm$^{-2}$ due
to the defect tolerance and defect reversing ability of the
amorphous structure and the larger band gap (1.8 eV). The carrier
mobility is very low ($\mu_e$ = 2-5 cm$^2$/Vs, $\mu_h$ = 0.005
cm$^2$/Vs), i.e. essentially only electrons contribute to the
signal. Basically any CMOS circuit and IC technology can be used
and technology changes are uncritical for this development. For
high-Z applications poly-crystalline HgI$_2$ constitutes a
possible semiconductor film material. The potential advantages are
small thickness, radiation hardness, and low cost. The development
is still in its beginnings and -- as for CMOS active pixel sensors
-- a real challenge to analog VLSI design.
\hfill \\
\item \emph{Amplification transistor implanted in high resistivity bulk} \\
In so-called DEPFET pixel sensors \cite{DEPFET-Lutz} a JFET or
MOSFET transistor is implanted in every pixel on a sidewards
depleted \cite{gatti84} bulk. Electrons generated by radiation in
the bulk are collected in a potential minimum underneath the
transistor channel thus modulating its current (fig. \ref{fig13}).
The bulk is fully depleted rendering large signals. The small
capacitance of the internal gate offers low noise operation. Both
together can be used to fabricate thin devices. The sensor
technology is non-standard and the operation of DEPFET pixel
detectors requires separate steering and amplification ICs (see
below).
\end{itemize}

\begin{figure}[h]
\begin{center}
\includegraphics[width=0.48\textwidth]{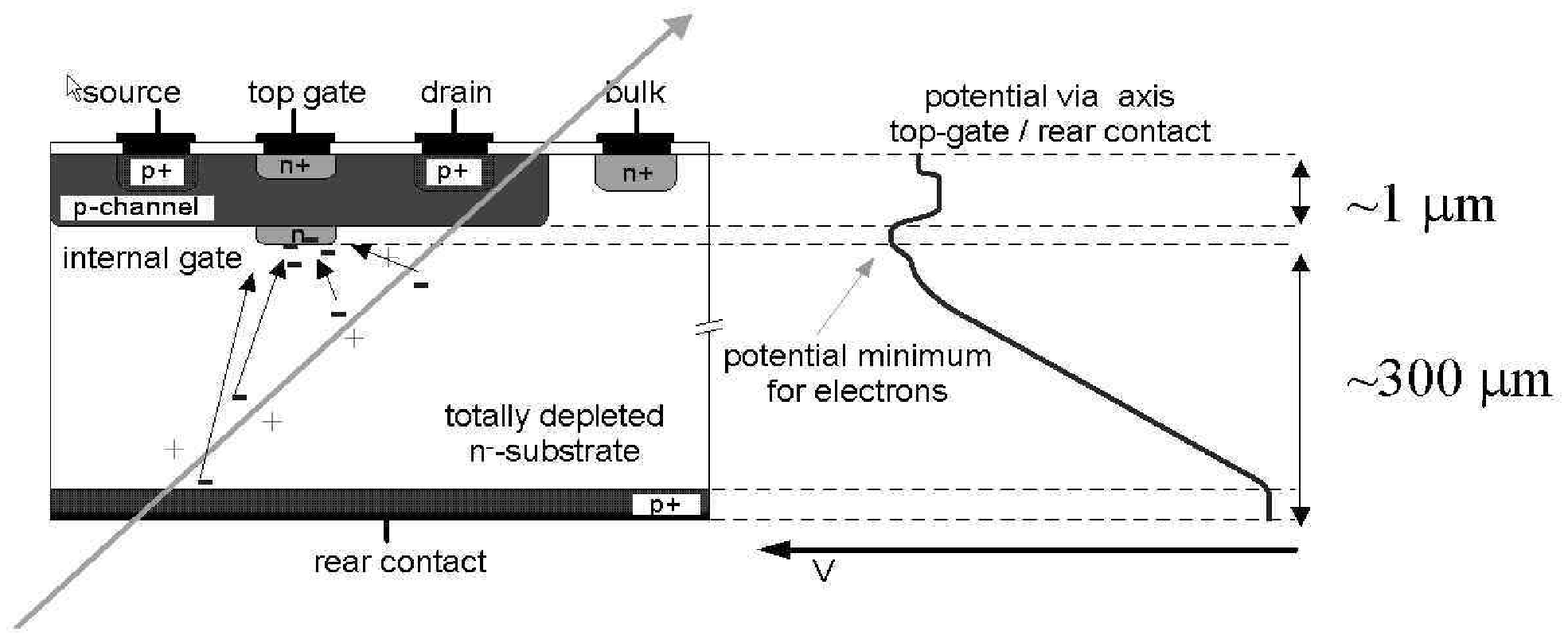}
\end{center}
\caption[]{Principle of operation of a DEPFET pixel structure
based on a sidewards depleted detector substrate material with an
imbedded planar field effect transistor. Cross section (left) of
half a pixel with symmetry axis at the
  left side, and potential profile (right).} \label{fig13}
\end{figure}

Because CMOS active pixels and DEPFET pixels are most advanced in
their development, they are discussed in more detail below.

\subsection{CMOS Active Pixels}
The main difference of CMOS monolithic active pixel sensors
\cite{meynants98} to CMOS camera chips is that they are larger in
area and must have a 100$\%$ fill factor for efficient particle
detection. Standard commercial CMOS technologies are used which
have a lightly doped epitaxial layer between the low resistivity
silicon bulk and the planar processing layer. The epi-layer is --
technology dependent -- at most 15$\mu$m thick and can also be
completely absent. Collaborating groups around IReS$\&$LEPSI
\cite{LEPSI,LEPSI-Portland}, RAL \cite{RAL-Vertex03} and
Irvine-LBNL-Ohio \cite{kleinfelder03} use similar approaches to
develop large scale CMOS active pixels also called MAPS
(Monolithic Active Pixel Sensors) \cite{MAPS}. Prototype detectors
have been produced in $0.6 \mu$m, $0.35 \mu$m and $0.25 \mu$m CMOS
technologies \cite{Dulinski03,AGay03}.

In the MAPS device (fig. \ref{fig11}(b)) electron-hole pairs
created by ionization energy loss in the epitaxial layer remain
trapped between potential barriers on both sides of the epi-layer
and reach, by thermal diffusion, an n-well/p-epi collection diode.
The sensor is depleted only directly under the n-well collection
diode where full charge collection is obtained; the charge
collection is incomplete in all other areas. With small pixel
cells collection times in the order of $100 ns$ are obtained. The
signal is small, some hundred to 1000e, depending on the thickness
of the epi-layer. A chip submission using a process with no
epi-layer, but with a low doped substrate of larger resistivity
has also been tried (MIMOSA-4) \cite{non-epi} which proved to
function with high detection efficiency for minimum ionizing
particles. This renders the uncertainty and dependence on the
future of the epi-layer thickness in different technologies less
constraining.

\begin{figure}[h]
\begin{center}
\includegraphics[width=0.20\textwidth]{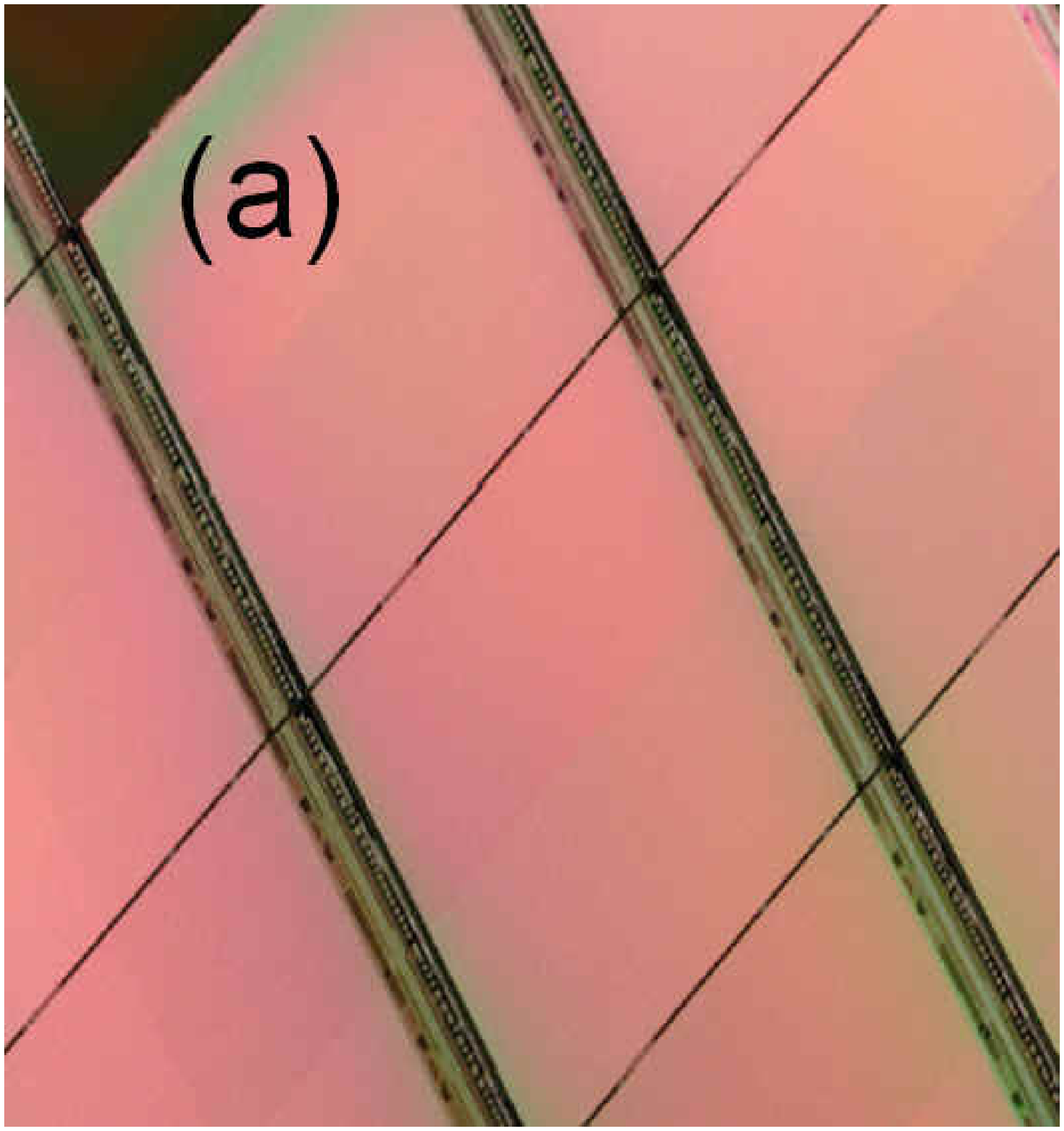}
\includegraphics[width=0.28\textwidth]{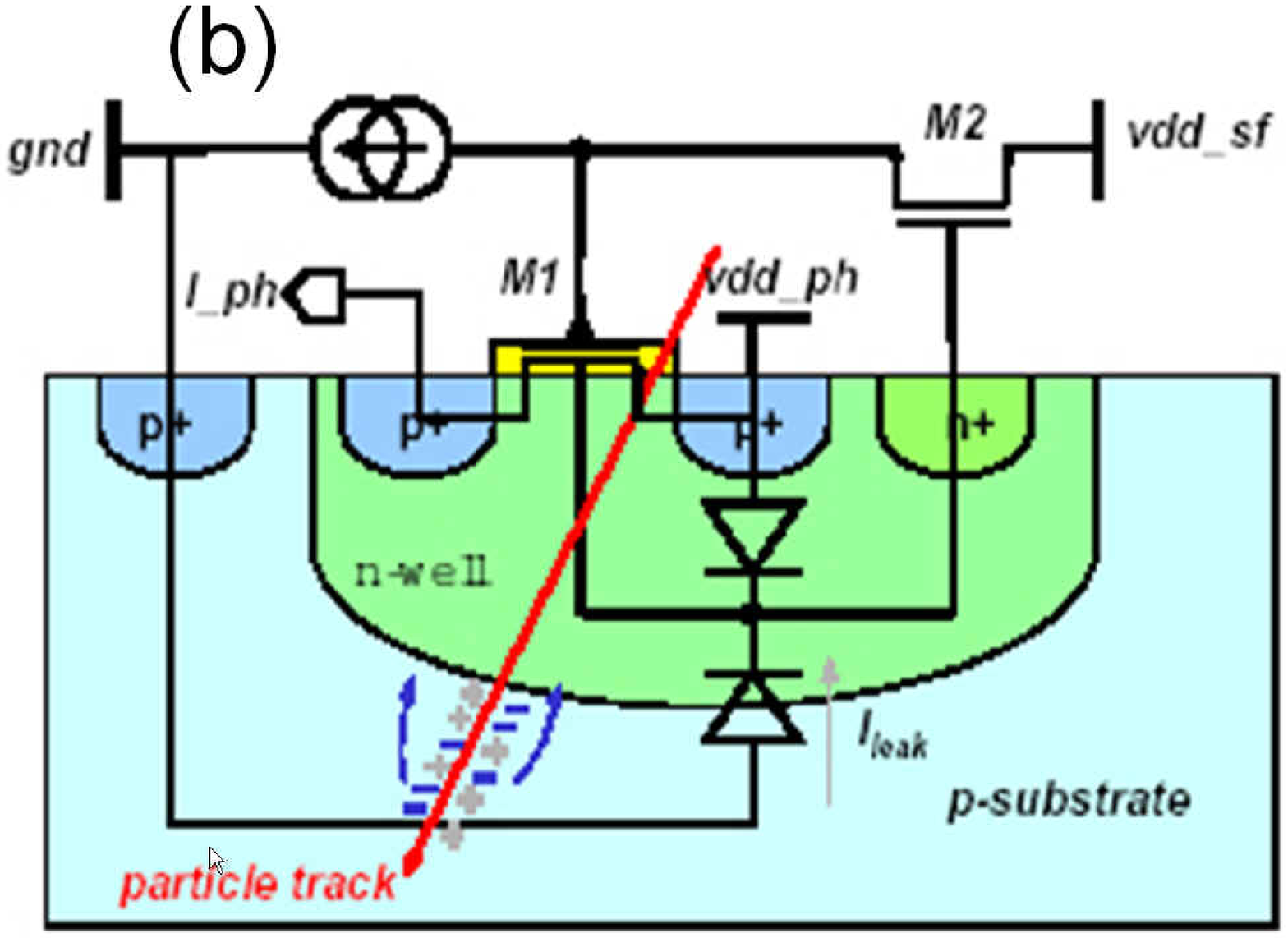}
\end{center}
\caption[]{(a) Seamless stitching in between wafer reticles to
obtain large area CMOS active pixels, (b) schematic of a pMOS
photo-FET to improve charge collection in MAPS.} \label{fig14}
\end{figure}

Matrix readout performed using a standard 3-transistor circuit
(line select, source-follower stage, reset) commonly employed by
CMOS matrix devices, but can also include current amplification
and current memory \cite{Dulinski03}. For an image two complete
frames are subtracted from each other (CDS) which suppresses
switching noise. Noise figures of 10-30e and S/N $\sim$20 have
been achieved with spatial resolutions below 5$\mu$m. The
radiation hardness of CMOS devices is always a crucial question
for particle detection in high intensity colliders. MAPS appear to
sustain non-ionizing radiation (NIEL) to $\sim$10$^{12}$n$_{eq}$
while the effects of ionizing radiation damage (IEL) are at
present still unclear \cite{dulinski}. Apart from this the focus
of further development lies in making larger area devices and in
increasing the charge collection performance in the epi-layer. For
the former, first results in reticle stitching, that is
electrically connecting IC area over the reticle border, have been
encouraging (fig. \ref{fig14}(a)) \cite{AGay03}. Full CMOS
stitching over the reticle boarder still has to be demonstrated.
The poor charge collection by thermal diffusion is another area
calling for ideas. Approaches using more than one n-well
collecting diode \cite{RAL-Vertex03}, a photo-FET
\cite{Dulinski03,LEPSI,LEPSI-Portland}, or a photo-GATE
\cite{Kleinfelder-Portland} are currently investigated. The
photo-FET called technique (fig. \ref{fig14}(b)) has features
similar to those of DEPFET pixels (see below). A pMOS FET is
implanted in the charge collecting n-well and signal charges
affect its gate voltage and modulate the transistor current.

Above all, the advantages of a fully CMOS monolithic device relate
to the adoption of standard VLSI technology and its resulting low
cost potential (potentially $\sim$25$\$$ per cm$^2$). In turn the
disadvantages also come largely from the dependence on commercial
standards. The thickness of the epi-layer varies for different
technologies. It becomes thinner for smaller processes and with it
also the number of produced signal electrons vanishes. Only a few
processing technologies are suited. With the rapid change of
commercial process technologies this is an issue of concern. The
fact that good detector performance has been seen with non-epi
wafers material provides some relief. Furthermore, in the active
area, due to the n-well collection diode, no CMOS (only nMOS)
circuitry is possible. The voltage signals are very small
($\sim$mV), of the same order as transistor threshold dispersions
requiring very low noise VLSI design. The radiation tolerance of
CMOS pixel detectors for particle detection beyond 10$^{12}$
n$_{eq}$ is still a problem although the achieved tolerance should
be sufficient for a Linear Collider. Improved readout concepts and
device development for high rate particle detection at a linear
collider is under development \cite{TESLA-TDR}. For the upgrade of
the STAR microvertex detector the use of CMOS active pixels is
planned \cite{Dulinski03}.

\subsection{DEPFET Pixels}
The Depleted Field Effect Transistor structure (DEPFET)
\cite{DEPFET-Lutz} provides detection and amplification properties
jointly and has been experimentally confirmed and successfully
operated as single pixel structures and as large (64x64) pixel
matrices \cite{DEPFET-Systems}. The DEPFET detector and implanted
amplification structure is shown in fig. \ref{fig13}. Sidewards
depletion \cite{gatti84} provides a parabolic potential which has
-- by appropriate biasing and a so-called deep-n implantation -- a
local minimum for electrons ($\sim 1 \mu$m) underneath the
transistor channel. The channel current can be steered and
modulated by the voltage at the external gate and - important for
the detector operation - also by the deep-n potential (internal
gate). Electrons collected in the internal gate are removed by a
clear pulse applied to a dedicated CLEAR contact outside the
transistor region or by other clear mechanisms. The very low input
capacitance ($\sim$ few fF) and the in situ amplification makes
DEPFET pixel detectors very attractive for low noise operation
\cite{ulrici03}. Amplification values of 400 pA per electron
collected in the internal gate have been achieved. Further
amplification enters at the second level stage (current based
readout chip).

\begin{figure}[htb]
\begin{center}
\includegraphics[width=.5\textwidth]{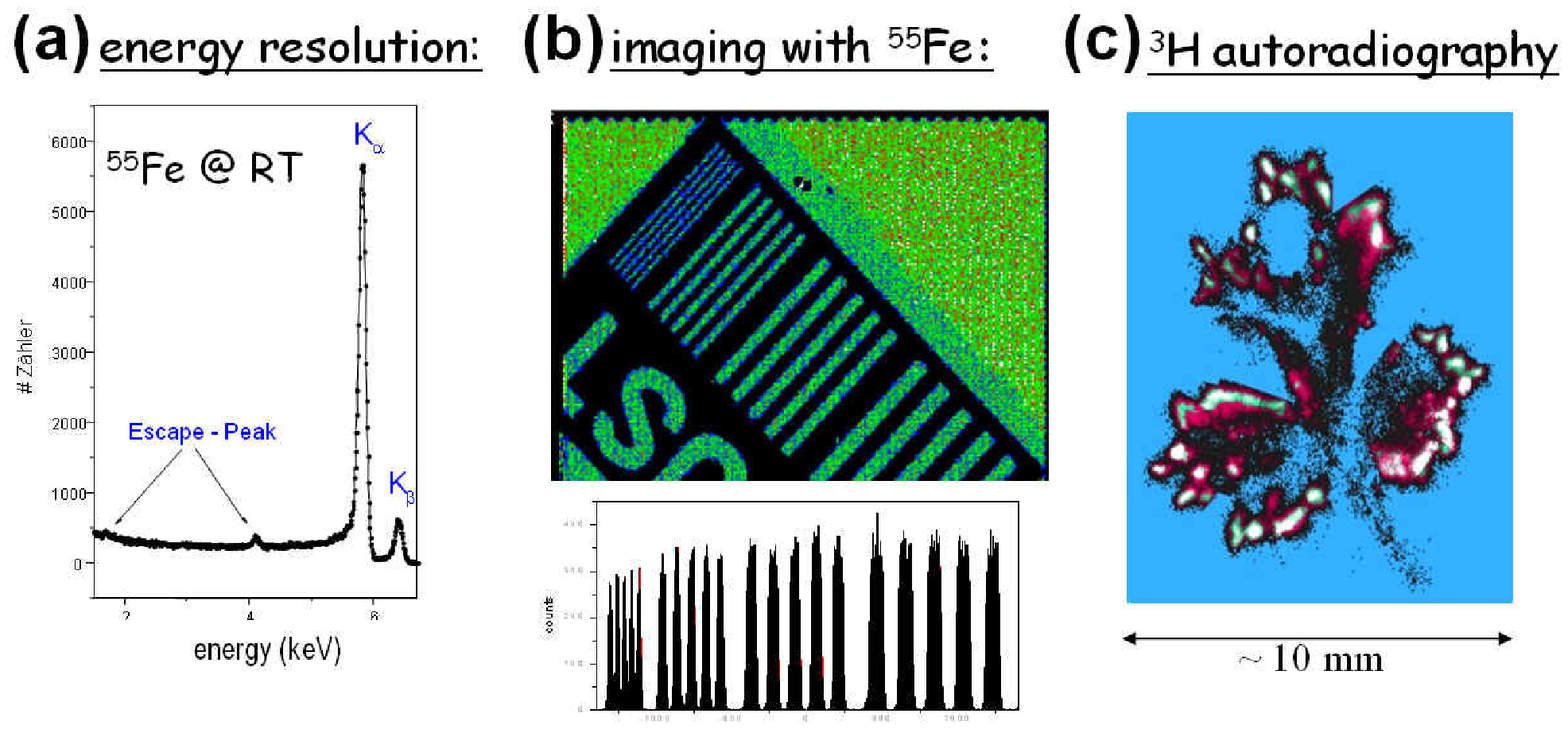}
\end{center}
\caption[]{(a) Response of a single DEPFET pixel structure to
$6$keV X-rays from an $^{55}$Fe source, (b) image obtained with
$^{55}$Fe and a test chart, the smallest structures have a pitch
of 50$\mu$m and are 25$\mu$m wide, (c) $^3$H labelled leaf.}
\label{fig15}
\end{figure}

DEPFET pixels are currently being developed for three very
different application areas: vertex detection in particle physics
\cite{DEPFET-TESLA,wermes-portland}, X-ray astronomy
\cite{DEPFET-XEUS} and for biomedical autoradiography
\cite{ulrici03}. Figure \ref{fig15} summarizes the achieved
performance in noise and energy resolution with single pixels
structures (fig. \ref{fig15}(a)) and in spatial resolutions with
64x64 pixels matrices (fig. \ref{fig15}(b)). With round single
pixel structures noise figures of 2.2e at room temperature and
energy resolutions of 131 keV for 6 keV X-rays have been obtained.
With small (20x30 $\mu$m$^2$) linear structures fabricated for
particle detection at a Linear Collider the noise figures are
about 10e. The spacial resolution read off from fig.
\ref{fig15}(b) in matrices operated with 50 kHz line rates is
\cite{wermes-portland}
\begin{eqnarray*}
\sigma_{xy} & = (4.3 \pm 0.8) \mu{\mathrm m} \quad {\mathrm or}
\quad 57 \frac{LP}{mm} \quad &{\mathrm {for \quad 22 \, keV}} \,
\gamma \\
\sigma_{xy} & = (6.7 \pm 0.7) \mu{\mathrm m} \quad {\mathrm or}
\quad 37 \frac{LP}{mm} \quad &{\mathrm {for \quad 6 \, keV}} \,
\gamma
\end{eqnarray*}

The capability to observe tritium in autoradiographical
applications is shown in fig. \ref{fig15}(c).

The very good noise capabilities of DEPFET pixels are very
important for low energy X-ray astronomy and for autoradiography
applications. For particle physics, where the signal charge is
large in comparison, this feature is used to design very thin
detectors ($\sim$50$\mu$m) with very low power consumption when
operated as a row-wise selected matrix \cite{wermes-portland}.
Thinning of sensors to a thickness of 50$\mu$m using a technology
based on deep anisotropic etching has been successfully
demonstrated \cite{laci-portland}. For the development of DEPFET
pixels for a Linear Collider matrix row rates of 50 MHz and frame
rates for 520x4000 pixels of 25 kHz, read out at two sides, are
targeted. Sensors with cell sizes of 20x30$\mu$m$^2$ have been
fabricated and complete clearing of the internal gate, which is
important for high speed on-chip pedestal subtraction, has been
demonstrated \cite{wermes-portland}. A large matrix is readout
using sequencer chips for row selection and current based chip for
column readout \cite{wermes-portland}. Both chips have been
developed at close to the desired speed for a Linear Collider. The
estimated power consumption for a five layer DEPFET pixel vertex
vertex detector is only 5W rendering a very low mass detector
without cooling pipes feasible. The most recent structures use
MOSFETs as DEPFET transistors for which the radiation tolerance
still has to be investigated.

\section{Summary}
Driven by the demands for high spatial resolution and high rate
particle detection in high energy physics, semiconductor pixel
detectors have started to also become exploited for imaging
applications. Hybrid pixel detectors, in which sensor and
electronic chip are separate entities, connected via bump bonding
techniques represent today's state of the art for both, particle
detection and imaging applications. New trends include interleaved
pixels having different pixel- and readout-pitches, MCM-D
structures and 3D-silicon detectors with active edges. Monolithic
or semi-monolithic detectors, in which detector and readout
ultimately are one entity, are currently developed in various
forms, largely driven by the needs for particle detection at
future colliders. CMOS active pixel sensors using standard
commercial technologies on low resistivity bulk and SOI pixels,
a-SI:H, and DEPFET-pixels, which try to maintain high bulk
resistivity for charge collection, classify the different ways to
monolithic detectors which are presently carried out.

\section*{Acknowledgment}
The author would like to thank W. Dulinski, M. Caccia, P. Jarron,
P. Russo, C. da Via', and S. Parker for providing new material and
information for this presentation.



%

\end{document}